\documentclass[aps,prl,twocolumn,preprintnumbers,amsmath,amssymb,superscriptaddress,showpacs]{revtex4-1}

\usepackage{graphicx}
\usepackage[T1]{fontenc}
\usepackage[colorlinks,bookmarks=false,citecolor=blue,linkcolor=blue,urlcolor=blue]{hyperref}
\usepackage[dvipsnames]{xcolor}
\usepackage{times}

\newcommand{\Eqref}[1]{Eq.~\eqref{#1}}

\begin{document}

\title{Subperiods and apparent pairing in integer quantum Hall interferometers}

\author{Giovanni A. Frigeri}
\affiliation{Institut f\"{u}r Theoretische Physik, Universit\"{a}t Leipzig, D-04103, Leipzig, Germany}
\affiliation{Max Planck Institute for Mathematics in the Sciences, D-04103, Leipzig, Germany}

\author{Daniel D. Scherer}
\affiliation{Niels Bohr Institute, University of Copenhagen, DK-2100 Copenhagen, Denmark}

\author{Bernd Rosenow}
\affiliation{Institut f\"{u}r Theoretische Physik, Universit\"{a}t Leipzig, D-04103, Leipzig, Germany}

\date{\today}

\begin{abstract}
We analyze the magnetic field and gate voltage dependence of the longitudinal resistance in an integer quantum Hall Fabry-P\'{e}rot interferometer, taking into account the interactions between an interfering edge mode, a non-interfering edge mode and the bulk. For weak bulk-edge coupling and sufficiently strong inter-edge interaction, we obtain that the interferometer operates in the Aharonov-Bohm regime with a flux periodicity halved with respect to the usual expectation. Even in the regime of strong bulk-edge coupling, this behavior can be observed as a subperiodicity of the interference signal in the Coulomb dominated regime. We do not find evidence for a connection between a reduced flux period and electron pairing, though. Our results can reproduce some recent experimental findings.
\end{abstract}


\maketitle

\begin{figure}[t!]
		\centering
		\includegraphics[width=\columnwidth]{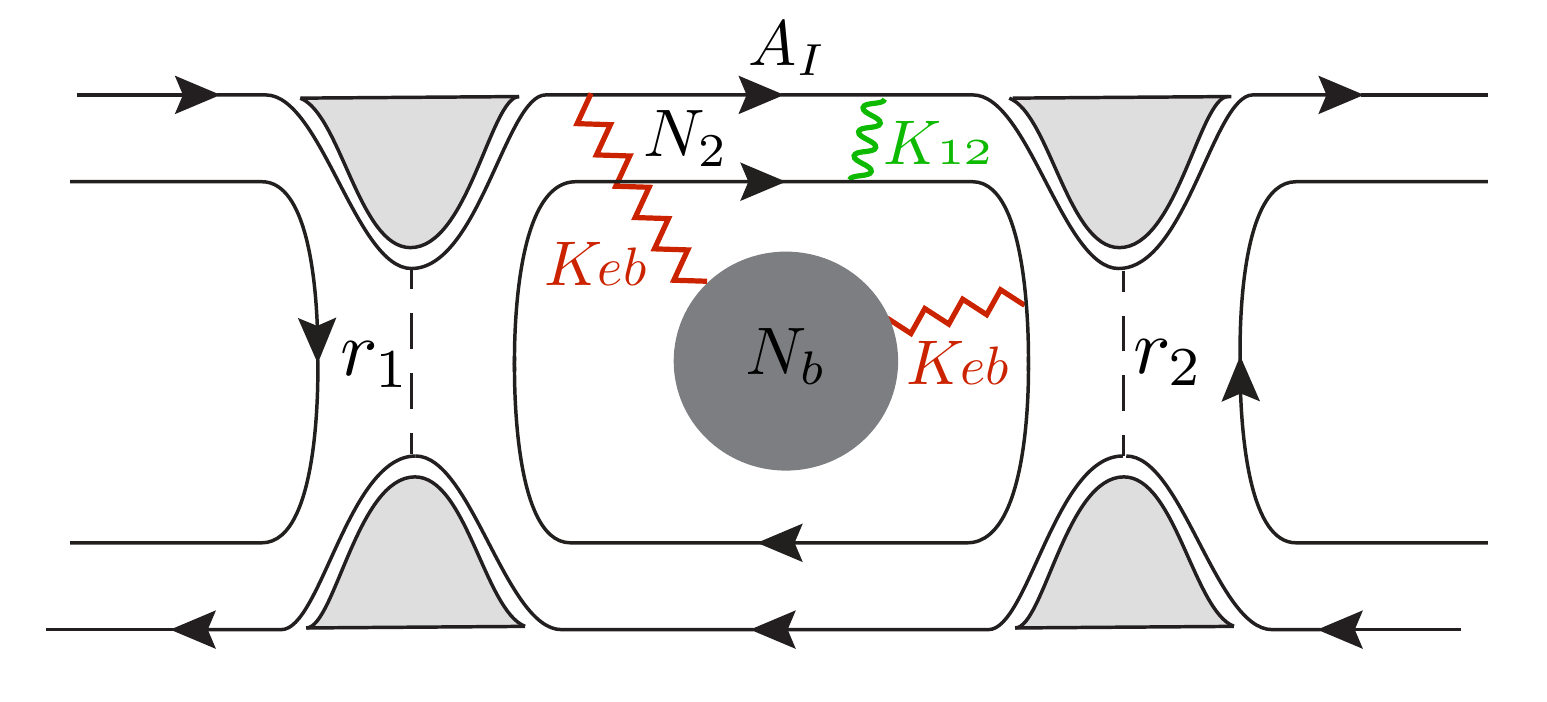}
		\caption{QH Fabry-P\'{e}rot interferometer in the open limit. The outermost interfering edge mode is coupled via Coulomb interactions to the second non-interfering edge mode (wavy line) as well as to the bulk of the system (zig-zag line).} \label{fig:FPI}
	\end{figure}

Phase coherence is a key ingredient of quantum mechanics, and its consequences can be observed in interference experiments. We consider here the electronic version of a Fabry-P\'{e}rot interferometer (FPI) realized in the integer quantum Hall (QH) regime~\cite{PhysRevB.55.2331,PhysRevB.72.155313,PhysRevB.76.155305,PhysRevLett.103.206806,zhang2009distinct,ofek2010role,choi2011aharonov,PhysRevLett.108.256804,choi2015robust,sivan2016observation}, consisting of a Hall bar perturbed by two constrictions, which introduce amplitudes for backscattering. The probability that a particle is backscattered is determined by the interference of trajectories with a phase difference given by the Aharonov-Bohm flux enclosed by the loop. Due to the flux-sensitivity of the phase difference, the backscattering probability oscillates as a function of the magnetic field $B$ and a gate voltage $V_G$ used to change the interferometer area $\bar{A}$. There has been a renewed interest in QH interferometers because they allow to reveal anyonic statistics~\cite{leinaas1977theory,wilczek1982magnetic,PhysRevLett.49.957,arovas1984fractional,stern2010non,PhysRevLett.96.016802,PhysRevLett.96.016803,Willett02062009,willett2010alternation,PhysRevLett.111.186401,PhysRevLett.115.126807} with possible applications to quantum computation~\cite{RevModPhys.80.1083,PhysRevLett.94.166802}.\\
Generally,  QH interferometers come in two variants: Aharonov-Bohm (AB) and Coulomb dominated (CD) ones~\cite{zhang2009distinct,ofek2010role,PhysRevLett.103.206806,choi2011aharonov,PhysRevLett.98.106801,PhysRevB.83.155440,PhysRevB.72.155313,PhysRevB.76.155305,PhysRevLett.108.256804,choi2015robust,sivan2016observation,PhysRevB.85.073403,NgoDinh20122794,baer2015transport}. The AB regime is characterized by a magnetic field periodicity  $\Delta B=\phi_0/\bar{A}$   and a gate voltage periodicity  $\Delta V_G= 1/\alpha$ \cite{PhysRevB.72.155313,zhang2009distinct}, where $\phi_0=h/|e|$ denotes    the flux quantum, 
and  $\alpha=(B/\phi_0)d\bar{A}/dV_G$. Moreover, the  lines of constant interference  phase in the two dimensional plane $V_G$-$B$ have negative slope~\cite{ofek2010role,zhang2009distinct}, because  the interference phase increases with both   magnetic field  and   interferometer area, 
and the latter grows when the gate voltage becomes more positive. One expects the AB signatures to be observed in large interferometers, when electrostatic interactions  are weak.  On the contrary, in small interferometers  bulk and edge are strongly coupled, placing the interferometer in the CD regime. In this case, the lines of constant phase have positive slope in the $V_G$-$B$ plane~\cite{sivan2016observation,ofek2010role,zhang2009distinct,PhysRevLett.108.256804}, and a reduced magnetic field period $\Delta B=\phi_0/(\nu \bar{A})$ is observed~\cite{PhysRevB.76.155305,choi2011aharonov,ofek2010role,zhang2009distinct}, with $\nu$ denoting the filling fraction in the constriction region. \\
Recently, halving of the magnetic field and gate voltage period was observed experimentally~\cite{choi2015robust} for a FPI with bulk filling factor $\nu_B$ between $2.5$ and $4.5$, i.e., $\Delta B=\phi_0/(2\bar{A})$ and $\Delta V_G=1/(2\alpha)$. Interestingly, the observed negative slope of lines with constant phase is consistent with  AB physics, while the reduced magnetic field period is reminiscent of CD physics. In addition, 
shot noise measurements yield a Fano factor of two, indicating that the halving of the flux period could be interpreted in terms of electron pairing~\cite{choi2015robust}. These intriguing experimental results are not explained by theoretical studies of Fabry-P\'{e}rot  interferometers so far~\cite{PhysRevLett.98.106801,PhysRevB.83.155440,PhysRevB.85.073403,NgoDinh20122794,2016arXiv160808889F,baer2015transport}.\\
In this Letter, we analyze a model which takes into account the electrostatic coupling of  two  modes as well as  bulk-edge coupling, appropriate for modeling the experiment Ref.~\onlinecite{choi2015robust}. We find that for  
strong inter-edge coupling and weak bulk-edge  interactions, the FPI is characterized by a negative slope of constant phase lines, together  with halved magnetic field and gate voltage periodicity, in agreement with the experimental results~\cite{choi2015robust}. We name this regime AB$^\prime$. When adding one flux quantum, one electron is added to each  edge mode, and due to the strong inter-edge coupling, this addition  occurs in an alternating fashion between outer and inner edge. When an electron is added to the inner edge, it induces a phase shift on the outer edge such that 
the oscillation period of the resistance is half a flux quantum. However, the analysis of  the  two-particle addition spectrum indicates that  electron pairing is unlikely to occur for a wide range of parameters.  We predict that an AB$^\prime$ subleading contribution is still present  even when the system is in the CD regime. Finally, we find that the transmission phase of a FPI embedded into a Mach-Zender interferometer (MZI) is described by AB physics, even for strong inter-edge coupling which places  the FPI in the AB$^\prime$ regime with regards to oscillations in the backscattering probability. \\
{\em Model \textemdash} We consider an electronic FPI at filling factor $2<\nu_B<3$ ($\nu_B\approx 3$) in the limit of weak backscattering, as depicted in Fig.~\ref{fig:FPI}. 
The outermost interfering edge encloses an area $A_I$, giving rise to an interference phase $\theta= 2 \pi A_I B/\phi_0$. We decompose the area $A_I = \bar{A}(V_G)+\delta A_I$, where  $\bar{A}$ varies slowly  with  gate voltage, while $\delta A_I$ represents a fluctuating part which is periodic in both magnetic field and gate voltage~\cite{PhysRevB.83.155440}.  The charges in the inner edge mode and in the bulk are quantized, and  are described by discrete variables $N_2, N_b \in \mathbb{Z}$.
%
%
Assuming that the area enclosed by the two edges and the bulk region are approximately equal, the charge imbalance in units of the electron charge is    $\delta Q_1= (A_I - \bar{A}) B/\phi_0$ on the interfering edge, $\delta Q_2=N_2- \bar{A}B/\phi_0 
-\phi_{02}$ on the non-interfering edge,  and  $\delta Q_b=N_b+2 \bar{A} B /\phi_0 - \overline{q}+ 2 \phi_{0b}$ in the bulk,  with $\overline{q}$ denoting the positive background charge, and with phase offsets  $\phi_{02}$ and $\phi_{0b}$. Accordingly,  the fluctuating part of the energy is given by
 %
	\begin{gather}\label{eq:energy}
		E= \frac{1}{2}\overrightarrow{\delta Q}^T \mathbb{V} \overrightarrow{\delta Q}\ , \\
			\overrightarrow{\delta Q}=
				\begin{pmatrix}
					\delta Q_1 \\ \delta Q_2 \\ \delta Q_b
				\end{pmatrix},\qquad
		\mathbb{V}=
				\begin{pmatrix}
					K_1    & K_{12} & K_{eb} \\
					K_{12} & K_2    & K_{eb} \\
					K_{eb} & K_{eb} & K_b
				\end{pmatrix} \nonumber,
	\end{gather}
%
where the diagonal terms in $\mathbb{V}$ represent the charging energies of the edges and the bulk, respectively. The off-diagonal terms encode the edge-edge coupling $K_{12}$ and the bulk-edge interaction $K_{eb}$. Considering all possible  winding numbers $m$ of the electron around the interference cell, the longitudinal resistance due to backscattering  is given by
%
	\begin{equation}\label{eq:backscattering probability}
		R\propto \left\langle\left|r_1+t_1 t_1^\prime\sum_{m=1}^{+\infty}(r_1^\prime)^{m-1} r_2^m  e^{im\theta}\right|^2\right\rangle,
	\end{equation}
%
where $\langle\dots\rangle$ represents a thermal average at inverse temperature $\beta$ over the $A_I, N_2,N_b$ fluctuations with respect to the energy function given in \Eqref{eq:energy}. Here, $r_i$, $t_i$ ($r^\prime_i$, $t^\prime_i$) denote reflection and transmission amplitudes at QPC $i$ for particles travelling along the upper (lower) edge. 
%

\begin{figure}[t!]
		\centering
		\includegraphics[width=0.7\columnwidth]{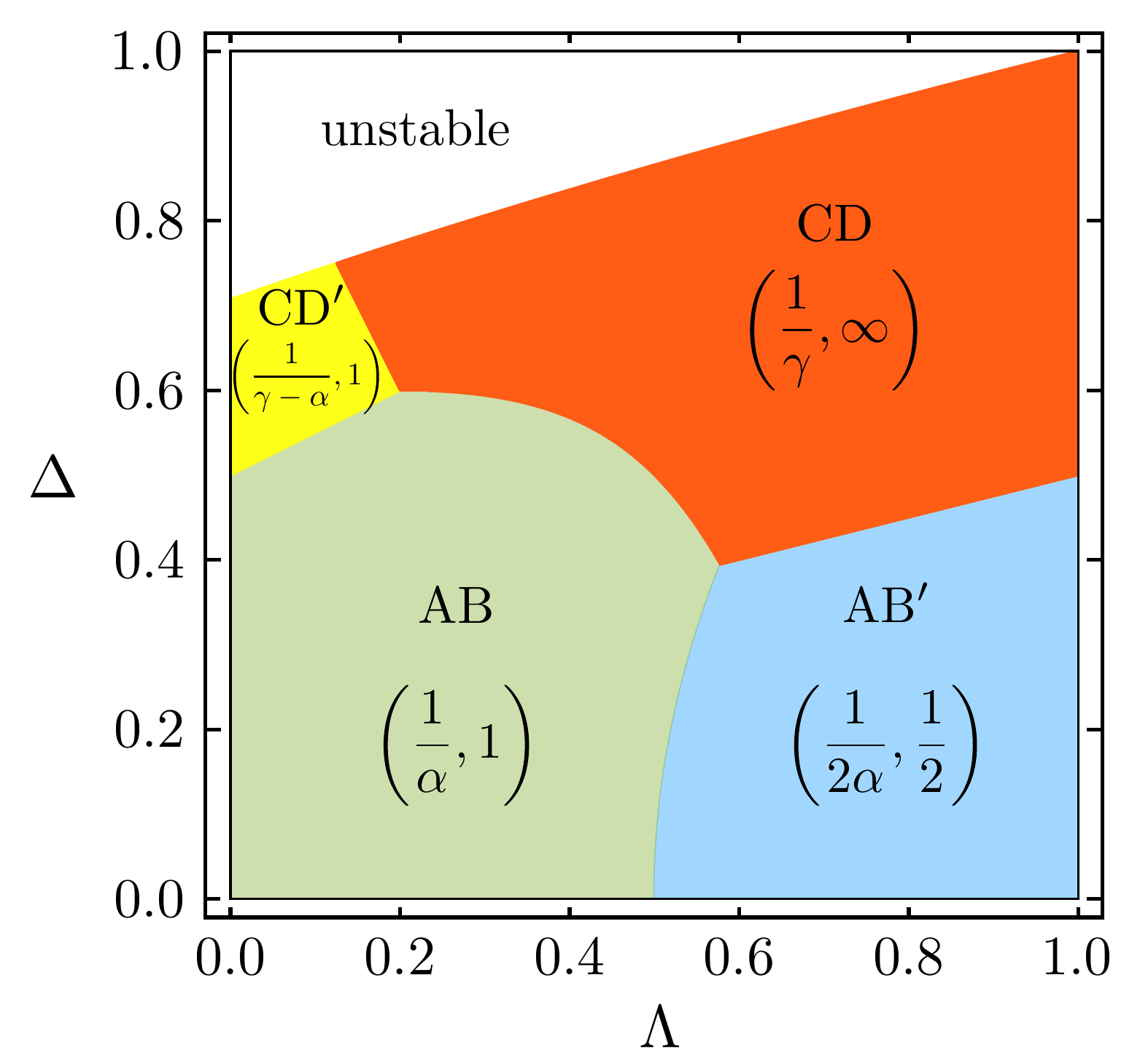}
		\caption{Phase diagram describing the leading term of \Eqref{eq:interference phase} as a function of $\Lambda=K_{12}/K_1$ and $\Delta=K_{eb}/K_1$ in the case of equal charging energies $K_1=K_2=K_b$. For each phase, gate voltage and magnetic field periods $(\Delta V_G,\Delta\phi)$ of the dominant term in $\langle e^{i\theta}\rangle$ are reported.}\label{fig:phasediagram}
	\end{figure}

\begin{figure}[t!]
		\centering
		\includegraphics[width=0.8\columnwidth]{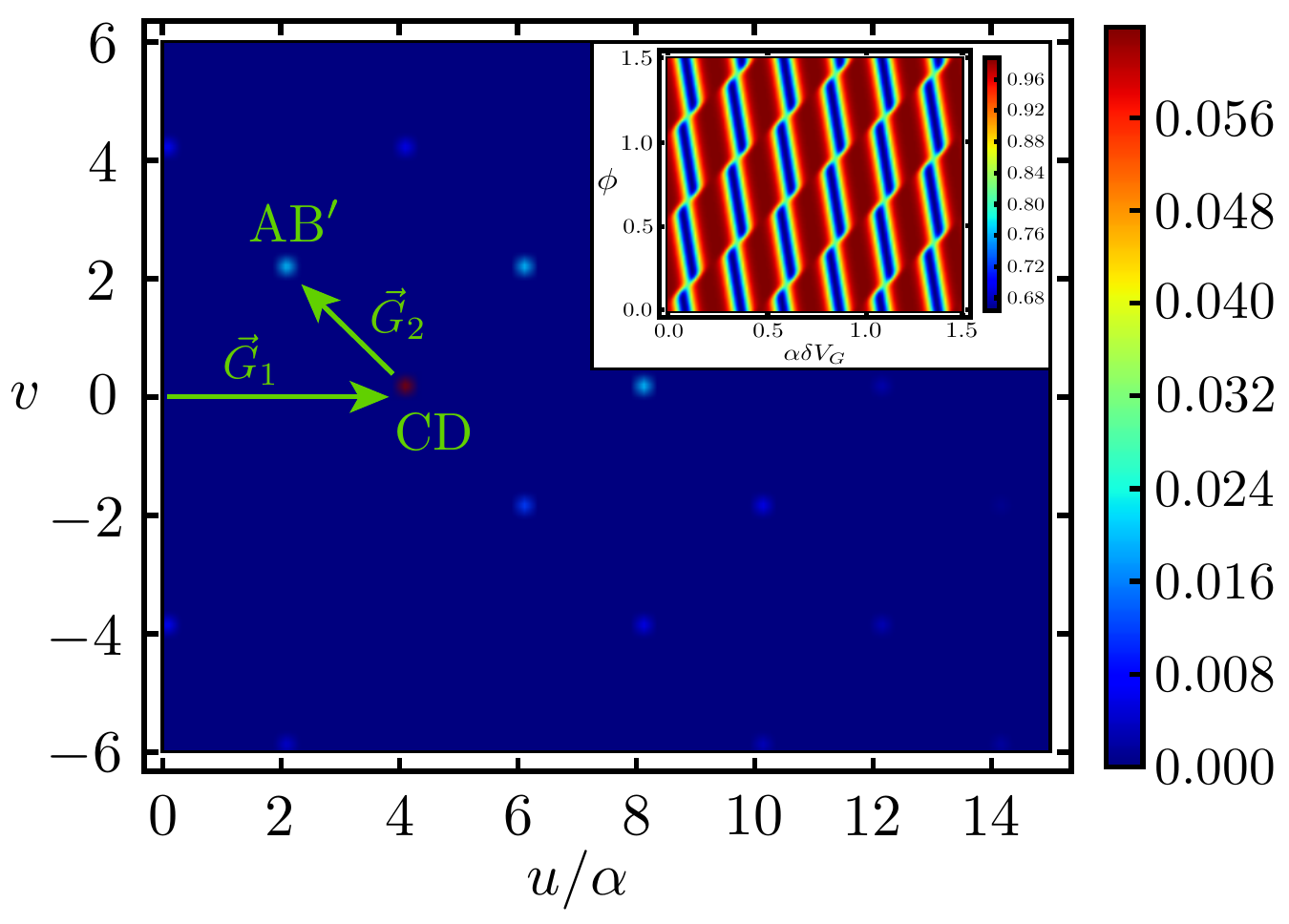}
		\caption{2D Fourier transform of the oscillatory part $\delta R$ of the longitudinal resistance~\Eqref{eq:backscattering probability} for $\Lambda=1$ and $\Delta=0.75$. The frequencies are spanned by reciprocal lattice vectors defined in \Eqref{eq:reciprocal lattice vectors}. The parameters are chosen as $r_1=r_2=0.9$, $\gamma=4\alpha$, $\phi_{02}=0.5$ and $\phi_{0b}=0.2$ at temperature $\zeta=0.01$. Inset: density plot of $R$ as a function of gate voltage and magnetic field.}\label{fig:fourier}
	\end{figure}

{\em Phase diagram \textemdash} We define the dimensionless parameters $\Delta=K_{eb}/K_1$, $\Lambda=K_{12}/K_1$ and $\zeta=(\beta K_1)^{-1}$. We let $K_1=K_2=K_b$  to simplify the discussion. These values are generic in the sense that the results will not change qualitatively when using different parameters. For this choice, the system is energetically stable if $|\Lambda|\leq 1$ and $|\Delta|\leq \sqrt{(1+\Lambda)/2}$ are satisfied. The expectation value of the interference phase $\langle e^{im\theta} \rangle$ in \Eqref{eq:backscattering probability} is evaluated by first minimizing the quadratic energy \Eqref{eq:energy} with respect to $A_I$.  To compute the remaining double sum over $N_2$ and $N_b$ we use the Poisson summation formula, and obtain %
	\begin{equation}\label{eq:interference phase}
		\frac{\left\langle e^{im\theta}\right\rangle}{e^{-2\pi^2m^2 \zeta}}=e^{2\pi m i\bar{A}B/\phi_0}
		\frac{\displaystyle\sum_{g,l}e^{-F(g+m\Delta,l+m\Lambda)}\mathcal{J}_{g,l}(\bar{A},B)}{\displaystyle\sum_{g,l}e^{-F(g,l)}\mathcal{J}_{g,l}(\bar{A},B)},
	\end{equation}
%
with $g$, $l\in\mathbb{Z}$, and the definitions
%
	\begin{align}
		&F(g,l):=\frac{2\pi^2 \zeta}{1-\Delta^2}\left\{g^2+\frac{\left[\Delta(1-\Lambda)g-(1-\Delta^2)l\right]^2}{(1-\Delta^2)(1-\Lambda^2)-\Delta^2(1-\Lambda)^2}\right\},\\
		&\mathcal{J}_{g,l}(\bar{A},B):=e^{2\pi i\left[g(2 \bar{A} B /\phi_0 +2\phi_{0b}  - \bar{q})-l(\bar{A} B/\phi_0+\phi_{02})\right]}.
	\end{align}
%
To leading order the denominator of~\Eqref{eq:interference phase} is unity, corresponding to $(g,l)=(0,0)$. 
We now define $  \gamma=d\bar{q}/dV_G $, and use the expansion 
%
\begin{equation}\label{eq:alpha and gamma}
 \bar{A} = \bar{A}_0 + \alpha \phi_0 \delta V_G/B   \ . 
\end{equation}
%
Introducing $\phi=\bar{A}_0 \delta B/\phi_0$, and considering the  numerator for
$m=1$, we find
%
	\begin{align}\label{eq:leading term}
		\left\langle e^{i\theta} \right\rangle &\approx w_{\text{AB}}e^{2\pi i(\phi+\alpha \delta V_G)}+w_{\text{AB$^\prime$}}e^{4\pi i(\phi+\alpha \delta V_G)}\nonumber\\
		&+w_{\text{CD}}e^{2\pi i \gamma \delta V_G}+w_{\text{CD$^\prime$}}e^{2\pi i[-\phi+(\gamma-\alpha) \delta V_G]},
	\end{align}
%
where the weight factors depend on the parameters $\Delta$, $\Lambda$ and also include  the phase offsets. In Fig.~\ref{fig:phasediagram} we indicate the regions in parameter space for which a given term in \Eqref{eq:leading term} is dominant.  
These regions are characterized  by the slope of lines with constant interference phase, the magnetic field period and the gate voltage period of the leading term of $\langle e^{i\theta} \rangle$. 
Interestingly, for strong edge-edge coupling and weak bulk-edge interactions the FPI is in the AB$^\prime$ regime (see Fig.~\ref{fig:phasediagram}), characterized by a negative slope of constant phase lines, 
together with halved flux and gate voltage periodicity. \\
%
%
{\em Subleading corrections in the CD regime\textemdash} Motivated by the experimental observation of a subleading 
AB component in a CD interferometer \cite{sivan2016observation}, we discuss the presence of a subleading AB$^\prime$ component for a CD interferometer with larger filling  $2 < \nu_B <3$ in the following. 
To be specific, we consider $\Lambda=1$ and $\Delta=0.75$, placing the interferometer in the CD regime as in~\cite{sivan2016observation}. Accordingly, we find from \Eqref{eq:interference phase}
%
	\begin{align}\label{eq:m=1}
		\left\langle e^{i\theta}\right\rangle_{\Lambda=1,\Delta=0.75}=&{\mathcal{A}}_1 e^{2\pi i \gamma \delta V_G}+{\mathcal{A}}_2 e^{4\pi i \left(\phi+\alpha \delta V_G\right)}\nonumber\\
		&+{\mathcal{A}}_3 e^{4\pi i \left[-\phi+(\gamma - \alpha)\delta V_G\right]}+\dots,
	\end{align}	
%
with the complex numerical coefficients ${\mathcal{A}}_j$ satisfying $|{\mathcal{A}}_j|>|{\mathcal{A}}_{j+1}|$ for $j=1,2,\dots$. We now consider higher winding numbers $m=2, 3$
%
	\begin{align}
		\left\langle e^{2i\theta}\right\rangle_{\Lambda=1,\Delta=0.75} =& {\mathcal{B}}_1 e^{4\pi i \gamma \delta V_G}+{\mathcal{B}}_2 e^{2\pi i \left[2\phi+(2 \alpha+\gamma)\delta V_G\right]}\nonumber\\
		&+{\mathcal{B}}_3 e^{2\pi i \left[-2\phi+(-2 \alpha+3\gamma) \delta V_G\right]}+\dots, \label{eq:m=2}\\
		\left\langle e^{3i\theta}\right\rangle_{\Lambda=1,\Delta=0.75} =& {\mathcal{C}}_1 e^{4\pi i \left[\phi+(\alpha+\gamma)\delta V_G\right]}+{\mathcal{C}}_2 e^{6\pi i \gamma \delta V_G}\nonumber\\
		&+{\mathcal{C}}_3 e^{2\pi i \left[4\phi+(4 \alpha+\gamma)\delta V_G\right]}+\dots,\label{eq:m=3}
	\end{align}
%
with the complex numerical coefficients $|\mathcal{{B}}_j|\geq|\mathcal{{B}}_{j+1}|$ and $|\mathcal{{C}}_j|>|\mathcal{{C}}_{j+1}|$. 
Contributions with even higher windings are suppressed by  powers  $(r_1^\prime)^{m-1} r_2^m$ and are thus less relevant. 
We see from \Eqref{eq:m=1} that the leading term of the interference phase is independent of magnetic field, 
and it has a gate voltage period $\Delta V_G= 1/\gamma$. 
Thus, the CD behavior dominates here and the lines of constant phase in the $V_G$-$\phi$ plane are mainly straight, as shown in  the inset of Fig.~\ref{fig:fourier}.  Nevertheless, 
also the AB$^\prime$ component and higher harmonics are revealed in the oscillations of $R$. \\
We note from~\Eqref{eq:m=1} that for fixed $m$ the subleading frequencies are shifted by $(\Delta V_G^{-1},\Delta\phi^{-1})=(2 \alpha - \gamma , 2)$ with respect to the dominant frequency. Moreover, 
multiple windings in Eqs.~(\ref{eq:m=2}), (\ref{eq:m=3}) translate a given frequency in \Eqref{eq:m=1} by integer multiples of $(\gamma,0)$. In order to formalize this observation, we introduce the 2D Fourier transform of $R$ as
%
	\begin{equation}\label{eq:Fourier}
		\tilde{R}(u,v)=\frac{\alpha}{N_\phi N_{\delta V_G}}\int\displaylimits_{0}^{N_\phi} \!  d\phi  \! \! \! \!
		\int\displaylimits_{0}^{\frac{N_{\delta V_G}}{\alpha}}   \! \! \! \! \!     d \delta V_G \,
		e^{-2\pi i (u \delta V_G+v \phi)} R(\delta V_G,\phi).
	\end{equation}
%
Thus, a specific interference pattern corresponds to a set of points in Fourier space $(u,v)$ described by 
 vectors $\vec{G}_{v_1,v_2}=v_1\vec{G}_{1}+v_2\vec{G}_{2}$, with $v_1\in\mathbb{Z}$ and different from zero, $v_2\in\mathbb{Z}$, and the basis  vectors  defined as
%
	\begin{equation}\label{eq:reciprocal lattice vectors}
		\vec{G}_{1}=\left(\gamma,0\right) \qquad\text{and}\qquad \vec{G}_{2}=\left(2 \alpha-\gamma,2\right).
	\end{equation}
%
%
%
In Fig.~\ref{fig:fourier} we show a density plot of $\tilde{R}(u,v)$ for $\Delta=0.75$, $\Lambda=1$, and compare its peaks to the vectors in~\Eqref{eq:reciprocal lattice vectors}. We note that if $\Lambda \neq 1$ also an additional subleading AB component would appear in the frequency spectrum of $R$.

\begin{figure}[t!]
		\includegraphics[width = \columnwidth]{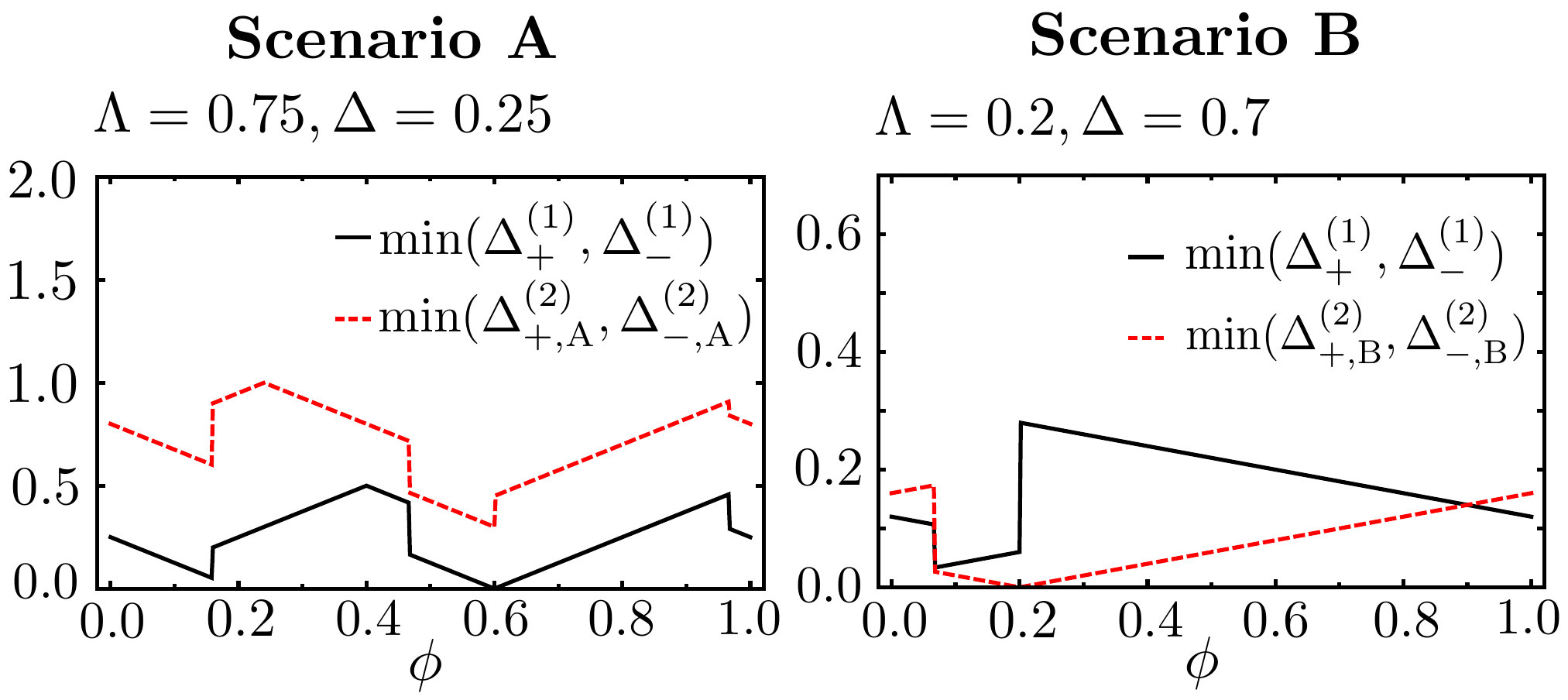}
		\caption{Minimum between addiction and subtraction energies as a function of $\phi$ for scenario A and B. There is no signature of electron pairing in scenario A for weak $K_{eb}$ and strong $K_{12}$. Pair tunneling can occur in scenario B if $K_{eb}>K_{12}$. We set $K_1 = K_2 = K_b$, $\delta V_G=0$,  $\phi_{02}=0.2$ and $\phi_{0b}=0.3$ in both cases.}\label{fig:addiction energies}
	\end{figure}

{\em Addition and subtraction energies \textemdash} 
The general idea of Coulomb mediated pairing is that two electrons are added to one subsystem, while expelling one electron from another subsystem, which may be  energetically favorable if 
the interaction within a given subsystem is weaker than the interaction between different subsystems \cite{Hamo+16}. 
To explore this possibility, we now study  a closed interferometer, 
which displays the same magnetic field and gate voltage periods as obtained by analyzing the open limit of the FPI 
\cite{Supplementary_material}.  
Conductance maxima are now related to degeneracies of the energy with respect to changing the number  $N_1$
of electrons on the outer edge, 
 since tunneling is assumed to occur only into the outermost edge (scenario A). 
Specifically, we compare the energy cost for adding, or subtracting, a single electron in the outermost edge with the corresponding energies for an electron pair in order to understand the transport processes occurring in the FPI. 
We obtain an energy function $E(N_1,N_2,N_b)$ from the one in~\Eqref{eq:energy} by setting $\delta Q_1 = N_1 - \bar{A}B/\phi_0$, and then define the addition and subtraction energies  $\Delta_\pm^{(1)}=E(N_1\pm1,N_2,N_b)-E(N_1,N_2,N_b)$, $\Delta_{\pm,\text{A}}^{(2)}=E(N_1\pm2,N_2 \mp 1,N_b)-E(N_1,N_2,N_b)$, allowing the number $N_2$ of electrons on the inner edge to relax in order to induce pairing. We plot both single particle and pair energies as a function of $\phi$ for weak $K_{eb}$ and strong $K_{12}$ (AB$^\prime$ phase). 
As can be seen from Fig.~\ref{fig:addiction energies}, adding or subtracting a pair of electrons needs more energy than in the case of a single electron. Therefore, we argue  that 
tunneling of electron pairs is generally not favored by energetic considerations, although there are special paramter values 
for which pair tunneling may be relevant \cite{Supplementary_material}. 
In principle, there is the possibility that correlated sequential tunneling may be important, but a rate equation analysis needed to answer this question is beyond the scope of the present manuscript.  \\
We next consider a scenario B, in which both inner and outer edge mode are contacted~\cite{baer2013cyclic} and we allow the bulk to relax due to the presence of the Ohmic contact in the center of the interferometer~\cite{choi2015robust}. We define the relaxed pair addition/subtraction energy $\Delta_{\pm,\text{B}}^{(2)}=E(N_1\pm1,N_2\pm1,N_b\mp1)-E(N_1,N_2,N_b)$. It is evident from Fig.~\ref{fig:addiction energies} that electron pair tunneling can now occur for some values of flux when $K_{eb}>K_{12}$. Here, ``pairing'' refers to a process in which the tunneling of one electron to the outer edge and a second one to the inner edge happens either simultaneously or strongly correlated in time.  
However, this scenario is not likely to be relevant for the experiment~\cite{choi2015robust}.
%
%

\begin{figure}[t]
		\centering
		\includegraphics[width=0.7\columnwidth]{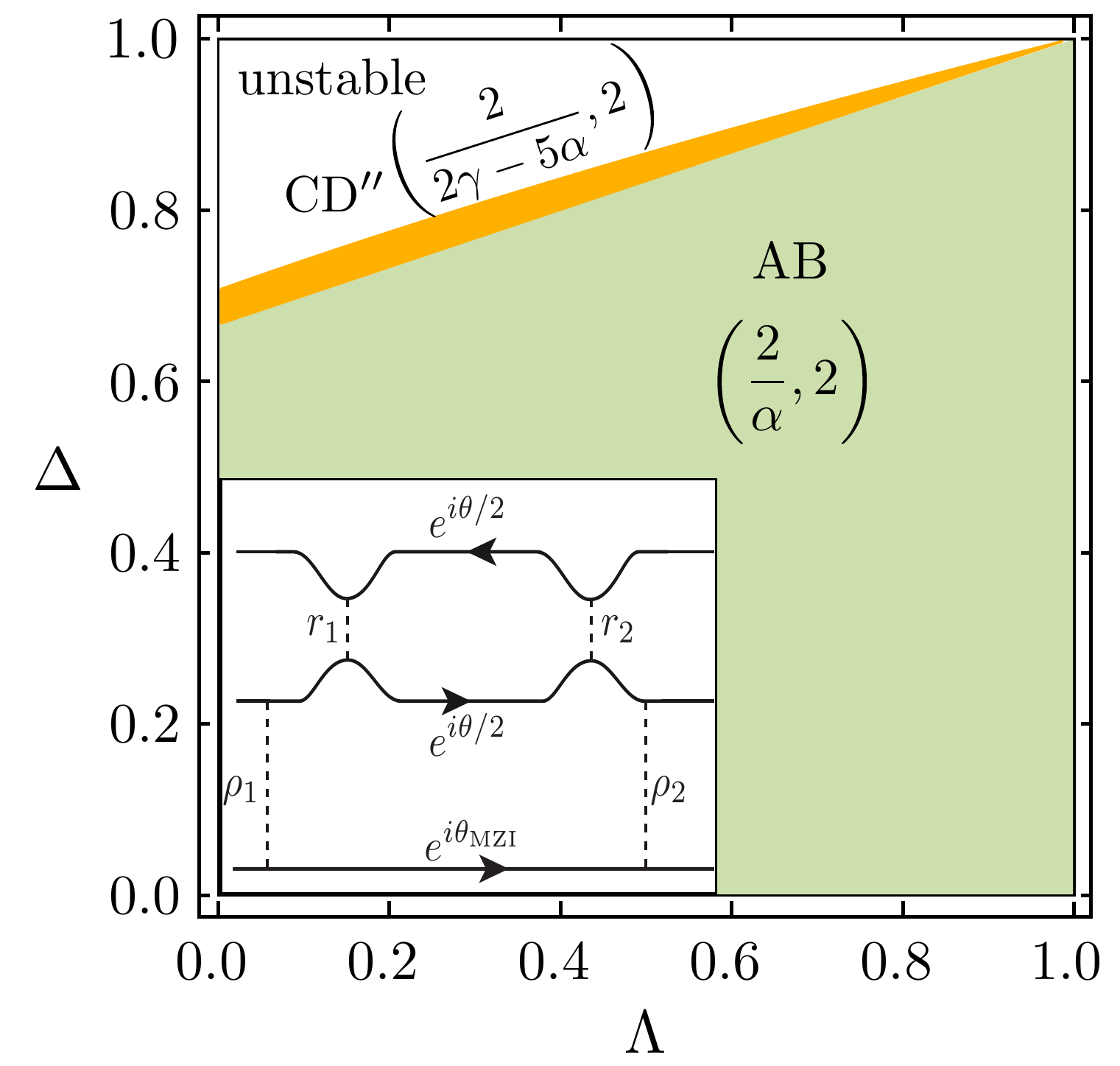}
	\caption{Phase diagram for the leading term of \Eqref{eq:leading-tau} as a function of $\Lambda=K_{12}/K_1$ and $\Delta=K_{eb}/K_1$ in the case $K_1=K_2=K_b$. For each phase the FPI gate voltage and magnetic field periodicity $(\Delta V_{G,\text{transm.}},\Delta\phi_\text{transm.})$ are reported. Inset: the MZI+FPI is reported at $\nu_B = 1$ for simplicity.\label{fig:transmission-leading}}
	\end{figure}

{\em Transmission phase \textemdash} We now consider a FPI placed in one of the two arms of a MZI \cite{sivan2017interaction}, which allows to measure the transmission phase through the FPI (see  inset of Fig.~\ref{fig:transmission-leading}). If the FPI is symmetric, then the Coulomb contribution to the phase is shared equally between upper and lower edge, and
we can express the MZI transmission probability as
%
	\begin{equation}\label{eq:tau}
		T_\text{MZI} = \left\langle \left| \tau_1 \tau_2 e^{i\theta_\text{MZI}}+\rho_1 \rho_2 t_1 t_2\frac{e^{\frac{i}{2}(\theta-2\pi\bar{A}B/\phi_0)}}{1-r_1^\prime r_2 e^{i\theta}}\right|^2\right\rangle,
	\end{equation}
%
where $\tau_1, \tau_2$ ($\rho_1, \rho_2$) are the transmission (reflection) amplitudes at the constrictions of the MZI and $\theta_\text{MZI} = 2\pi A_\text{MZI} B /\phi_0$.  The subtraction of $2 \pi \bar{A} B/\phi_0$ is done in order to avoid double counting  the FPI contribution to the MZI interference phase. Setting $m=1/2$ in \Eqref{eq:interference phase}, we obtain that
%
	\begin{align}\label{eq:leading-tau}
		\left\langle e^{i\theta/2} \right\rangle \approx & \omega_\text{AB} e^{i\pi (\phi+\alpha \delta V_G)} \left[1+\eta e^{2\pi i(\phi+\alpha \delta V_G)}\right] \nonumber\\
		&+ \omega_\text{CD''} e^{i\pi \left[-\phi+\left(2\gamma- 5\alpha\right)\delta V_G\right]},
	\end{align}
%
with $\eta\ll 1$ and the weight factors $\omega$ depending on the values of $\Delta$ and $\Lambda$. We indicate in Fig.~\ref{fig:transmission-leading} the $(\Lambda,\Delta)$ regions for which a given term in \Eqref{eq:leading-tau} dominates.
%
%
One sees that the phase diagram is almost completely composed of AB phase, where the FPI gate voltage and magnetic field periods are trivially doubled with respect to the AB periods of the longitudinal resistance $R$ due to the choice of $m=1/2$ (see Fig.~\ref{fig:phasediagram}). This behavior is independent of the edge-edge interaction strength $\Lambda$. Only for large 
$\Delta$,  a small region with CD features emerges. From Eqs. (\ref{eq:tau})-(\ref{eq:leading-tau}) we see that  $T_\text{MZI}$ has the leading periodicity $\Delta B = \phi_0/A_\text{MZI}$, and  a subleading component with  $\Delta B = \phi_0/(A_\text{MZI}-\bar{A}_0)$,  even when $r_1 = r_2 = 0$ such that interfering electrons do not encircle the interferometer cell. 
These findings are in agreement with the recent experiment~\cite{sivan2017interaction}.


{\em Comparison with experiments \textemdash} In the experiment~\cite{choi2015robust} the FPI was found to be in the AB regime with magnetic field periodicity $\Delta B_\text{exp.}^{\nu_B=2} = \phi_0/(12.5 \mu m^2)$ and gate voltage periodicity $\Delta V_{G,\text{exp.}}^{\nu_B=2} = \phi_0 /(0.17 B \mu m^2 V^{-1})$.  At $\nu_B\simeq 3$ (representative of $2.5<\nu_B<4.5$) however, the periodicities are $\Delta B_\text{exp.}^{\nu_B=3} = \phi_0/(25 \mu m^2)$ and $\Delta V_{G,\text{exp.}}^{\nu_B=3} = \phi_0 /(0.34 B \mu m^2 V^{-1})$, with the lines of constant phase  still having negative slope. The AB phase at  $\nu_B\simeq 2$ is theoretically described by the limit $\Lambda =0$,  and according to Fig.~\ref{fig:phasediagram} we find $\Delta B_\text{theo.}^{\nu_B=2} = \phi_0/\bar{A}_0$ and $\Delta V_{G,\text{theo.}}^{\nu_B=2} = 1/\alpha$. Assuming that at $\nu_B\simeq 3$ the FPI is in the AB$^\prime$ phase, the periodicities are $\Delta B_\text{theo.}^{\nu_B=3} = \phi_0/(2\bar{A}_0)$ and $\Delta V_{G,\text{theo.}}^{\nu_B=3} = 1/2 \alpha$. Therefore,  the experimentally
observed periodicities Ref.~\onlinecite{choi2015robust} are reproduced by our model. A discussion of  current noise~\cite{choi2015robust} and interference with multiple areas~\cite{sivan2017interaction} are beyond the scope of the present manuscript.\\
Experimentally, 
fingerprints of the AB phase in a CD-interferometer were reported in Ref.~\onlinecite{sivan2016observation}, and fingerprints of 
the AB$^\prime$ phase were observed  at weaker magnetic field ~\cite{Private_communication}.
In particular,  the dominant CD component had a gate voltage periodicity $\Delta V_\text{CD,exp.}^{\nu_B=2}= 1/222 V$, and the subleading AB period was $\Delta V_\text{AB,exp.}^{\nu_B=2}= 1/90 V$ at $B=5T$ ($1<\nu_B<2$) . According to our model, $\Delta V_\text{AB,theo.}^{\nu_B=2} = 1/\alpha$ and $\Delta V_\text{CD,theo.}^{\nu_B=2}=1/\gamma$  at $1<\nu_B<2$~\cite{Supplementary_material}, yielding parameters   $\alpha_{B=5T}= 90 V^{-1}$ and $\gamma_{B=5T}=222V^{-1}$. Since $\alpha$ depends linearly on the applied magnetic field $B$ and $\gamma$ is independent of it, 
for a weaker magnetic field $B=3.3 T$  ($2<\nu_B<3$), we 
expect a leading CD gate voltage period $\Delta V_\text{CD,theo.}^{\nu_B=3} = 1/\gamma = 1/222 V$ and a subleading AB$^\prime$ period $\Delta V_\text{AB$^\prime$,theo.}^{\nu_B=3} = 1/(2\alpha) = 1/118.8 V$.   These  results are in agreement with recent experimental findings~\cite{Private_communication}. \\
{\em Conclusion \textemdash} We propose a theoretical model for a FPI in which the outermost interfering edge interacts with the second non-interfering edge as well as with the bulk of the system. We find that for weak bulk-edge and strong edge-edge coupling the resistance oscillations are of AB type with halved flux periodicity (AB$^\prime$ regime), in agreement with recent experimental results~\cite{choi2015robust}. However, we do not find evidence for the importance of electron pair tunneling. We argue  that there are fingerprints of AB$^\prime$ physics even if the system is in the CD regime thanks to a partial screening of the bulk-edge interactions. We also calculate the transmission phase for a FPI situated in one arm of a MZI. 

\emph{Acknowledgements} ---
We would like to thank M.~Heiblum and I.~Sivan  for helpful discussions and for sharing unpublished results,
and the German Science Foundation (DFG) for financial support via grant RO 2247/8-1.

%

\pagebreak
\widetext
\begin{center}
	\textbf{\large Supplementary material for "Subperiods and apparent pairing in integer quantum Hall interferometers"}
\end{center}
\setcounter{equation}{0}
\setcounter{figure}{0}
\setcounter{table}{0}
\setcounter{page}{1}
\makeatletter
\renewcommand{\theequation}{S\arabic{equation}}
\renewcommand{\thefigure}{S\arabic{figure}}
\renewcommand{\bibnumfmt}[1]{[S#1]}
\renewcommand{\citenumfont}[1]{S#1}
\section{\label{sec:Calculation of the interference phase} S1. Calculation of the interference phase}

We provide here some details about the calculation of the interference phase $\langle e^{im\theta}\rangle$. We define the dimensionless parameters $\bar{K}_2=K_2/K_1$, $\bar{K}_b=K_b/K_1$, $\zeta = (\beta K_1)^{-1}$ where $\beta$ in the inverse temperature . In the open limit, we can evaluate the thermal expectation value
%
\begin{align}\label{eq:s-interference phase}
\langle e^{im\theta}\rangle &= \frac{1}{\mathcal{Z}} \sum_{N_2,N_b}\int_{-\infty}^{+\infty}dA_I \exp\left[-\beta E+2\pi m i A_IB/\phi_0\right]
= \frac{1}{\mathcal{Z}}\sqrt{2\pi \zeta} \exp\left[-2\pi^2 m^2 \zeta+2\pi m i\bar{A}B/\phi_0\right] \nonumber\\
&\times\sum_{N_2,N_b}\exp\Bigg\{-\frac{1}{2\zeta}\left[\left(\bar{K}_b-\Delta^2\right)(N_b+2\bar{A}B/\phi_0+2\phi_{0b}-\bar{q})^2
+\left(\bar{K}_2-\Lambda^2\right)(N_2-\bar{A}B/\phi_0-\phi_{02})^2\right] \nonumber\\
&\qquad\qquad\qquad\qquad -2\pi m i\left[\Delta(N_b+2\bar{A}B/\phi_0+2\phi_{0b}-\bar{q})+\Lambda(N_2-\bar{A}B/\phi_0-\phi_{02})\right] \nonumber\\
&\qquad\qquad\qquad\qquad -\frac{\Delta}{\zeta}\left(1-\Lambda\right)(N_2-\bar{A}B/\phi_0-\phi_{02})(N_b+2\bar{A}B/\phi_0+2\phi_{0b}-\bar{q})\Bigg\},
\end{align}
%
where the partition function $\mathcal{Z}$ is
%
\begin{align}\label{eq:partition function}
\mathcal{Z} &=\sum_{N_2,N_b}\int_{-\infty}^{+\infty}dA_I \exp\left[-\beta E\right] \nonumber\\
&=\sqrt{2\pi \zeta} \sum_{N_2,N_b}\exp\Bigg\{-\frac{1}{2\zeta}\left[\left(\bar{K}_b-\Delta^2\right)(N_b+2\bar{A}B/\phi_0+2\phi_{0b}-\bar{q})^2+
\left(\bar{K}_2-\Lambda^2\right)(N_2-\bar{A}B/\phi_0-\phi_{02})^2\right] \nonumber\\
&\qquad\qquad\qquad\qquad\qquad -\frac{\Delta}{\zeta}\left(1-\Lambda\right)(N_2-\bar{A}B/\phi_0-\phi_{02})(N_b+2\bar{A}B/\phi_0+2\phi_{0b}-\bar{q})\Bigg\}.
\end{align}
%
We note that the energy $E$ defined in the main text is quadratic in the variable $A_I$, hence to obtain \Eqref{eq:s-interference phase} and \Eqref{eq:partition function} we just solved a Gaussian integral.
%
%
The summations over the discrete degrees of freedom $N_2$, $N_b$ can be managed through the Poisson summation formula %
\begin{equation}\label{eq:Poisson}
\sum_{N=-\infty}^{+\infty}=\int_{-\infty}^{+\infty}dN \sum_{g=-\infty}^{+\infty}e^{-2\pi i N g} \ \  .
\end{equation}
%
Hence, \Eqref{eq:Poisson} allows to integrate over $N_{2}$ $N_b$ instead of summing, and we   obtain that
%
\begin{gather}
\frac{\langle e^{im\theta} \rangle}{e^{-2\pi^2 m^2 \zeta}} =  \frac{\mathcal{N}_m}{\mathcal{Z}} := e^{2\pi m i \bar{A}B/\phi_0} \frac{\displaystyle{\sum_{g,l} e^{ -F(g+m\Delta,l+m\Lambda)+2\pi i\left[g(2\bar{A}B/\phi_0+2\phi_{0b}-\bar{q})-l(\bar{A}B/\phi_0+\phi_{02})\right]}}}{\displaystyle{\sum_{g,l} e^{ -F(g,l)+2\pi i\left[g(2\bar{A}B/\phi_0+2\phi_{0b}-\bar{q})-l(\bar{A}B/\phi_0+\phi_{02})\right]}}} \label{eq:oscilllation two edges}
\end{gather}
%
with $g,l\in\mathbb{Z}$ and the function $F(g,l)$ defined as
%
\begin{equation}\label{eq:F different coupling}
F(g,l) := \frac{2\pi^2\zeta}{\bar{K}_b-\Delta^2}\left\{g^2+\frac{\left[\Delta(1-\Lambda)g-(\bar{K}_b-\Delta^2)l\right]^2}{(\bar{K}_b-\Delta^2)(\bar{K_2}-\Lambda^2)-\Delta^2(1-\Lambda)^2}\right\}.
\end{equation}
%
If we assume equal charging energies for the bulk and the edges, i.e., $\bar{K}_2=\bar{K}_b=1$, \Eqref{eq:F different coupling} coincide with Eq. (5) in the main text. First of all, we consider the leading order of the \Eqref{eq:oscilllation two edges}. We focus only on the numerator because to leading order the denominator is equal to unity, since its dominant term given by the duple $(g,l)=(0,0)$. The integers $g$ and $l$ that give rise to the dominant term in the numerator depend on the value of $\Lambda$ and $\Delta$. We note that varying $\bar{K}_2$ and $\bar{K}_b$ does not affect the phase diagram in a qualitative way;   moreover the phase diagram is independent of the parameters $\alpha, \gamma, \phi_{02}, \phi_{0b}$ and $\zeta$. The AB$^\prime$ phase is extended by increasing $\bar{K}_2$, while the CD phases expand with an increase of $\bar{K}_b$. The interesting AB$^\prime$ phase always appears in the phase diagram if $\bar{K}_2>1/4$.
\section{\label{sec:Subleading corrections to the interference phase} S2. Subleading corrections to the interference phase}
In the following we set $\bar{K}_2=\bar{K}_b=1$ and consider the limit of  strong edge-edge coupling, i.e. $\Lambda=1$. Therefore, \Eqref{eq:oscilllation two edges} simplifies to
%
\begin{gather}
\frac{\langle e^{im\theta}\rangle}{e^{-2\pi^2 m^2 \zeta}}\Bigg|_{\bar{K}_2=\bar{K}_b=1, \Lambda=1}=\frac{\mathfrak{N}_m}{\mathfrak{Z}} := e^{2\pi m i(2\bar{A}B/\phi_0+\phi_{02})}\frac{\displaystyle{\sum_{g=-\infty}^{+\infty} \exp\left[-\bar{F}(g+m\Delta)+2\pi i g(2\bar{A}B/\phi_0+2\phi_{0b}-\bar{q})\right]}}{\displaystyle{ \sum_{g=-\infty}^{+\infty} \exp\left[-\bar{F}(g)+2\pi i g(2\bar{A}B/\phi_0+2\phi_{0b}-\bar{q})\right]}}, \label{eq:oscillation K12=K}
\end{gather}
%
with the function $\bar{F}$ defined as 
%
\begin{equation}\label{eq:F}
\bar{F}(g):= \left(\frac{2\pi^2 \zeta}{1-\Delta^2}\right)g^2,
\end{equation}
%
because we need to keep only the term $l=-m$ in the numerator $\mathcal{N}_m$ of \Eqref{eq:oscilllation two edges} and $l=0$ in the denominator $\mathcal{Z}$. The leading term of the denominator $\mathfrak{Z}$ in \Eqref{eq:oscillation K12=K} is given by $g=0$, while its $n$-th sub-leading term is a result of the $g=\pm n$ terms. From these considerations and from the Taylor expansion of $(1+x)^{-1}\simeq 1-x+x^2$, we obtain that
%
\begin{align}\label{eq:denominator}
\mathfrak{Z}^{-1}&\simeq(1+2 e^{-\bar{F}(1)}\cos\left[2\pi(2\bar{A}B/\phi_0+2\phi_{0b}-\bar{q})\right]+2 e^{-\bar{F}(2)}\cos\left[4\pi(2\bar{A}B/\phi_0+2\phi_{0b}-\bar{q})\right]+\dots)^{-1} \nonumber\\
&\simeq 1-2 e^{-\bar{F}(1)}\cos\left[2\pi(2\bar{A}B/\phi_0+2\phi_{0b}-\bar{q})\right]+4 e^{-2\bar{F}(1)}\cos^2\left[2\pi(2\bar{A}B/\phi_0+2\phi_{0b}-\bar{q})\right]+\dots
\end{align}
%
We note that this result is valid for any value of the edge-bulk coupling $\Delta$. For $\Delta=0.75$ the leading behavior and subleading corrections of the numerator $\mathfrak{N}_m$ in \Eqref{eq:oscillation K12=K} are
%
%
%
\begin{align}
\mathfrak{N}_{m=1} \simeq & e^{-\bar{F}(-1/4)} \exp\left[2\pi i \left(\bar{q}+\phi_{02}-2\phi_{0b}\right)\right]+e^{-\bar{F}(3/4)}\exp\left[2\pi i(2\bar{A}B/\phi_0+\phi_{02})\right] \nonumber\\
&+e^{-\bar{F}(-5/4)}\exp\left[2\pi i\left(-2\bar{A}B/\phi_0+2\bar{q}-4\phi_{0b}+\phi_{02}\right)\right]+\dots \label{eq:N1}\\
\mathfrak{N}_{m=2} \simeq & e^{-\bar{F}(-1/2)}\exp\left[4\pi i \left(\bar{q}+\phi_{02}-2\phi_{0b}\right)\right]+e^{-\bar{F}(-1/2)}\exp\left[2\pi i(2\bar{A}B/\phi_0+\bar{q}+\phi_{02}-2\phi_{0b})\right] \nonumber\\
&+e^{-\bar{F}(-3/2)}\exp\left[2\pi i\left(-2\bar{A}B/\phi_0+3\bar{q}-6\phi_{0b}+2\phi_{02}\right)\right]+e^{-\bar{F}(-3/2)}\exp\left[4\pi i\left(2\bar{A}B/\phi_0+\phi_{02}\right)\right]+\dots \label{eq:N2}\\
\mathfrak{N}_{m=3} \simeq & e^{-\bar{F}(1/4)} \exp\left[2\pi i \left(2\bar{A}B/\phi_0+2\bar{q}+3\phi_{02}-4\phi_{0b}\right)\right]+e^{-\bar{F}(-3/4)}\exp\left[6\pi i(\bar{q}+\phi_{02}-2\phi_{0b})\right] \nonumber\\
&+e^{-\bar{F}(5/4)}\exp\left[2\pi i\left(4\bar{A}B/\phi_0+\bar{q}-2\phi_{0b}+3\phi_{02}\right)\right]+\dots 	\label{eq:N3}
\end{align}
%
Considering small variation of gate voltage $\delta V_G = V_G-V_G^{(0)}$ and magnetic field $\delta B = B - B^{(0)}$ from an initial value $V_G^{(0)}, B^{(0)}$, we have $\bar{A} = \bar{A}_0+\alpha\phi_0 \delta V_G/B$ and $\bar{q} = \bar{q}_0+\gamma \delta V_G$. Defining $\phi = \bar{A}_0 \delta B/\phi_0$ and combining \Eqref{eq:denominator} with Eqs.~(\ref{eq:N1})-(\ref{eq:N2})-(\ref{eq:N3}), we obtain the Eqs.~(8)-(9)-(10) in the main text.
\section{\label{sec:Interferometer with only one edge} S3. Interferometer at $1<\nu_B<2 (\nu_B\approx 2$)}

In the case the FPI consists of a single interfering edge and the bulk, the energy function $E(A_I,N_b)$ is
%
\begin{equation}
E(A_I,N_b) = \frac{K_1}{2} \left[(A_I-\bar{A})B/\phi_0\right]^2+\frac{K_b}{2} (N_b+\bar{A}B/\phi_0+\phi_{0b}-\bar{q})^2+K_{eb} \left[(A_I-\bar{A})B/\phi_0\right] (N_b+\bar{A}B/\phi_0+\phi_{0b}-\bar{q}).
\end{equation}
%
Proceeding as before, we obtain in this case that
%
\begin{equation}\label{eq:final-oscillation}
\displaystyle
\frac{\langle e^{im\theta}\rangle}{e^{-2\pi^2m^2\zeta}} = \frac{\mathcal{\tilde{N}}_m}{\mathcal{\tilde{Z}}} :=e^{2\pi m i \bar{A}B/\phi_0}\times\frac{\displaystyle\sum_{g=-\infty}^{+\infty}\exp\left[-\bar{F}(g+m\Delta)+2\pi i g(\bar{A}B/\phi_0+\phi_{0b}-\bar{q})\right]}{\displaystyle\sum_{g=-\infty}^{+\infty}\exp\left[-\bar{F}(g)+2\pi i g(\bar{A}B/\phi_0+\phi_{0b}-\bar{q})\right]},
\end{equation}
%
with $\bar{F}(g)$ given by \Eqref{eq:F}. The dominant term in the denominator $\mathcal{\tilde{Z}}$ is $g=0$; on the other hand the leading term of the numerator $\mathcal{\tilde{N}}$ depends on the value of $\Delta$~\cite{s-PhysRevB.83.155440}. The $n$-th subleading term of the denominator $\mathcal{\tilde{Z}}$ is given by $g=\pm n$ terms. Therefore, we obtain from the Taylor expansion that
%
\begin{equation}\label{eq:1/D}
\mathcal{\tilde{Z}}^{-1}\simeq 1-2 e^{-\bar{F}(1)}\cos\left[2\pi(\bar{A}B/\phi_0+\phi_{0b}-\bar{q})\right]+4e^{-2\bar{F}(1)}\cos^2\left[2\pi(\bar{A}B/\phi_0+\phi_{0b}-\bar{q})\right]+\dots
\end{equation}
%
We now set $\Delta=0.75$; our choice is motivated by the experimental setup in \cite{s-sivan2016observation}. In this case we can write the numerator as
%
\begin{align}
\mathcal{\tilde{N}}_{m=1}\big|_{\Delta=0.75}\simeq & e^{-\bar{F}(-1/4)} \exp\left[ 2\pi i \left(\bar{q}-\phi_{0b}\right)\right]+e^{-\bar{F}(3/4)}\exp(2\pi i \bar{A}B/\phi_0)+\dots \label{eq:N m=1}\\
\mathcal{\tilde{N}}_{m=2}\big|_{\Delta=0.75}\simeq & e^{-\bar{F}(1/2)}\exp\left(4\pi i \bar{q}-\phi_{0b}\right) + e^{-\bar{F}(1/2)} \exp\left[ 2\pi i \left(\bar{A}B/\phi_0+\bar{q}-\phi_{0b}\right)\right]+\dots \label{eq:N m=2}
\end{align}
%
Combining \Eqref{eq:1/D} with Eqs.~(\ref{eq:N m=1})-(\ref{eq:N m=2}), the oscillatory part of the resistance in terms of $\delta V_G$ and $\phi$ is
%
\begin{align}
\langle e^{i\theta}\rangle\big|_{\Delta=0.75} = &\mathcal{\tilde{A}}_1 \exp\left( 2\pi i \gamma \delta V_G\right)+\mathcal{\tilde{A}}_2 \exp\left[2\pi i (\phi+\alpha \delta V_G)\right] +\mathcal{\tilde{A}}_3 \exp\left\{ 2\pi i \left[-\phi+(2\gamma-\alpha)\delta V_G\right]\right\} \label{eq:s-m=1}\\ 
\langle e^{2i\theta}\rangle\big|_{\Delta=0.75} = &\mathcal{\tilde{B}}_1 \exp\left( 4\pi i \gamma \delta V_G\right)+\mathcal{\tilde{B}}_2 \exp\left\{ 2\pi i \left[\phi+(\alpha+\gamma) \delta V_G\right]\right\} +\mathcal{\tilde{B}}_3 \exp\left\{ 2\pi i \left[-\phi+(3\gamma-\alpha)\delta V_G\right]\right\} \label{eq:s-m=2}
\end{align}
%
where $|\mathcal{\tilde{A}}_i|>|\mathcal{\tilde{A}}_{i+1}|$ and $|\mathcal{\tilde{B}}_i|\geq|\mathcal{\tilde{B}}_{i+1}|$, with $i=1,2,\dots$.
\begin{figure}[t]
	\begin{minipage}{0.45\textwidth}
		\centering
		\includegraphics[scale=0.5]{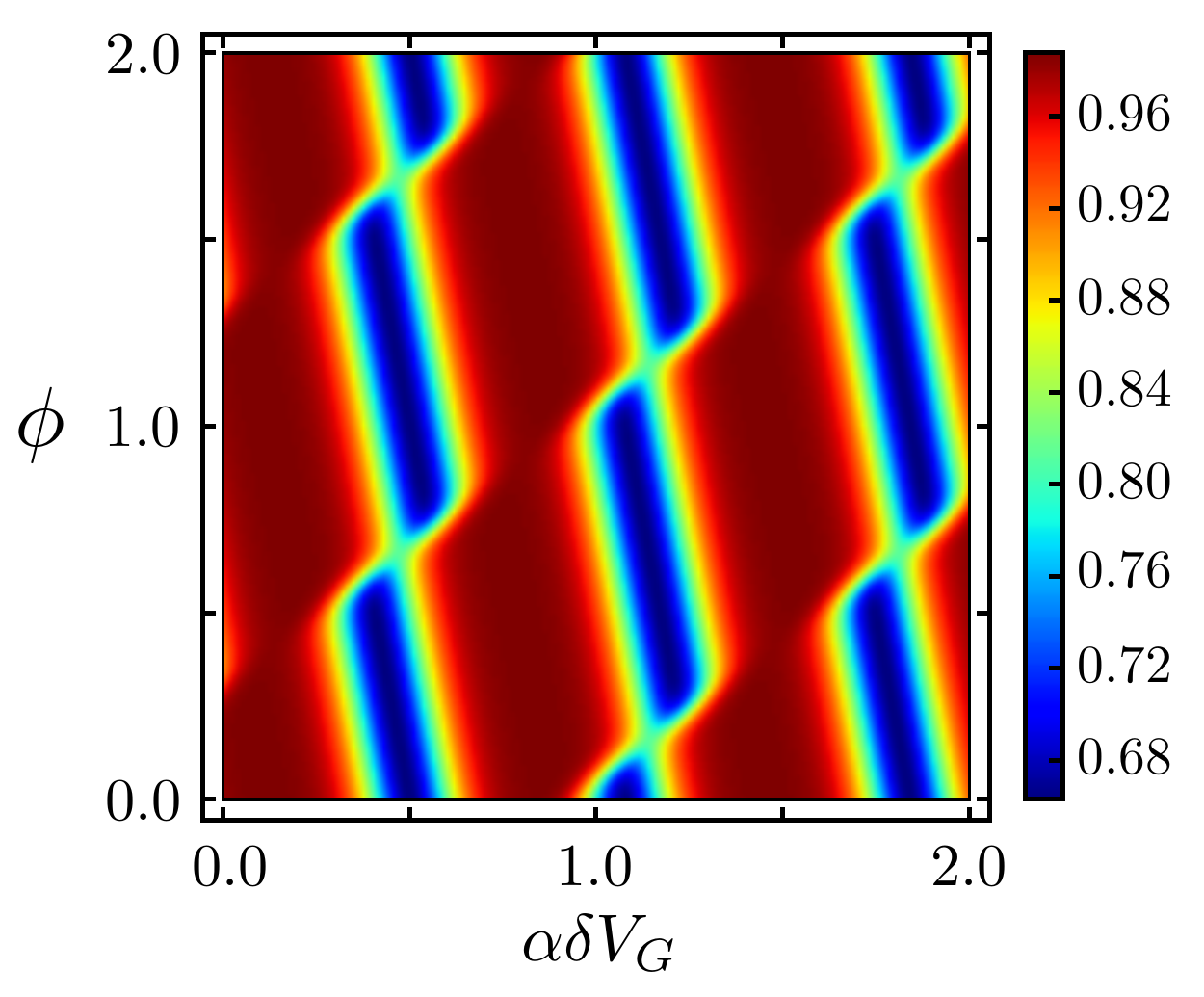}
	\end{minipage}
	\begin{minipage}{0.45\textwidth}
		\centering
		\includegraphics[scale=0.5]{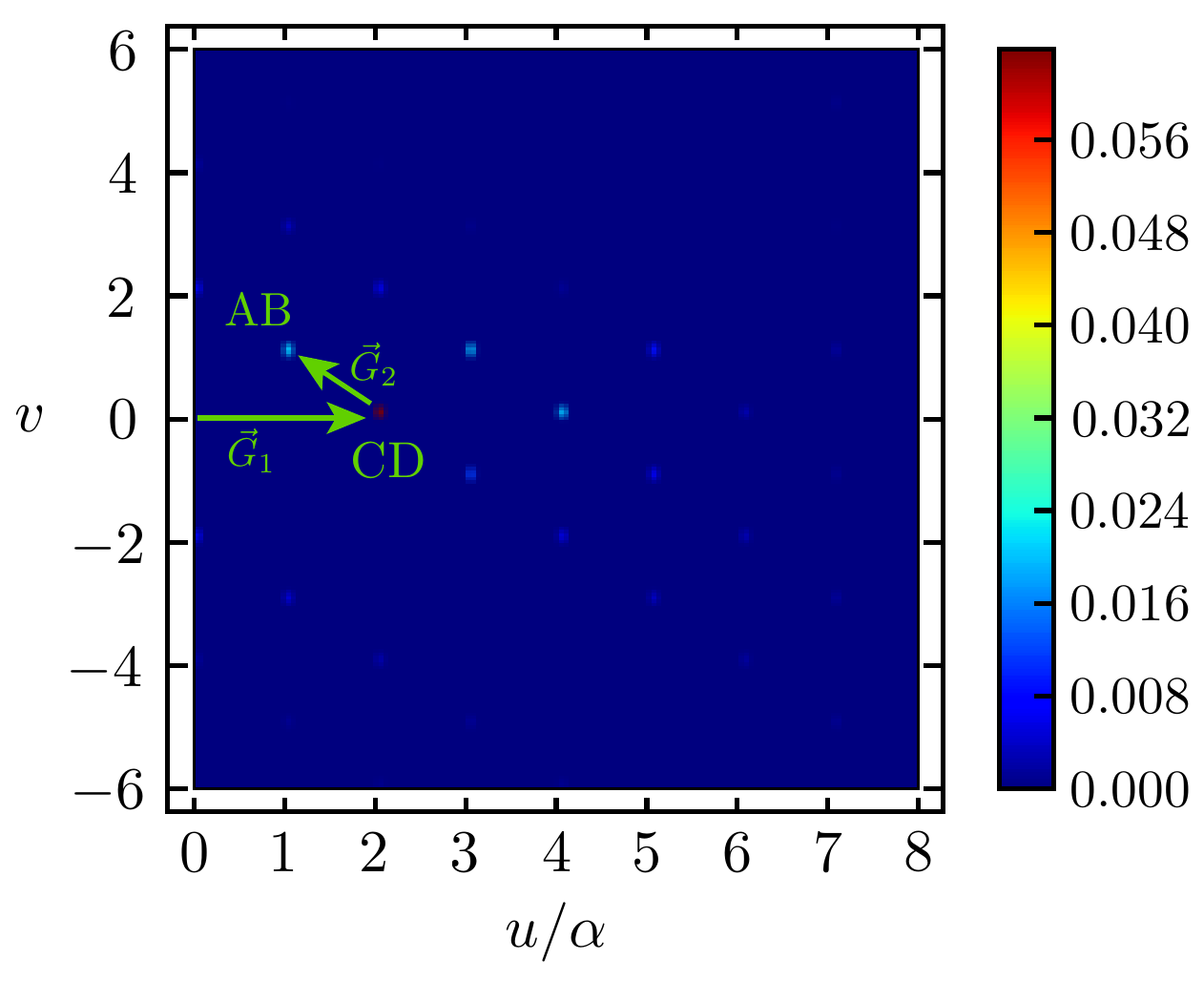}
	\end{minipage}
	\caption{Density plot of the backscattering probability $R$ at $1 < \nu_B < 2$ for $\Delta = 0.75$ and the corresponding 2D Fourier transform. The parameters are chosen as $r_1=r_2=0.9$, $\gamma=2\alpha$ and $\phi_{0b}=0.2$ at temperature $\zeta=0.01$.}
	\label{fig:s-fourier}
\end{figure}
Therefore, thanks to a partial screening of the edge-bulk interaction, the resistance displays also AB component and higher harmonics other than the dominant CD term.
From Eqs.~(\ref{eq:s-m=1})-(\ref{eq:s-m=2}), we can identify the peaks of  the 2D Fourier transform of $R$ with  vectors $\vec{G}_{v_1,v_2}=v_1\vec{G}_{1}+v_2\vec{G}_{2}$ (see Fig.\ref{fig:s-fourier}), with $v_1\in\mathbb{Z}$ and different from zero, $v_2\in\mathbb{Z}$, and with the two reciprocal lattice vectors now defined as
%
\begin{equation}\label{eq:s-reciprocal lattice vectors}
\vec{G}_{1}=\left(\gamma,0\right) \qquad\text{and}\qquad \vec{G}_{2}=\left(\alpha-\gamma,1\right).
\end{equation}
%
Fourier components $\vec{G}_{0,v_2}$ show up in the experimental results~\cite{s-sivan2016observation}, while these are missing in our theoretical analysis so far. The lack of the terms with $v_1=0$ is caused by our assumption that the backscattering amplitudes $r_{1,2}$ are independent of the charge in the bulk. If we assume a linear dependence of $r_i\simeq r_{i,0}+c N_b$, we find that the additional frequency $\vec{G}_{0,v_2}$ arise from the direct term $\langle|r_i|^2\rangle$.\\ 
\section{\label{sec:Pairing scenarios} S4. Pairing scenarios}
We now consider the model in the closed limit, with $N_1\in\mathbb{Z}$ representing the quantized charge in the interfering edge mode.
Despite its simplicity, our model in principle allows for electron pairing to occur, in case the mutual Coulomb repulsion energetically favors such processes with changes of $\Delta N_i =2$ in one of the edges. We have identified two distinct mechanisms  that feature a sequence of transitions in the groundstate stability diagram of the closed system relevant for pairing. The first case corresponds to scenario A of the main text, and is realized for $K_{2} > K_{12} > K_{1}$ and $K_{12}$ in the vicinity of the unstable point of the quadratic form~Eq.(1) in the main text. Without loss of generality, we neglect the bulk and set $K_{eb} = 0$. The system first undergoes a flux-induced charge re-distribution as $(N_{1},N_{2}) \rightarrow (N_{1}-1,N_{2}+1)$ and as the flux is further increased, a $\Delta N_{1} = 2$ transition followed by a redistribution $(N_{1}-1,N_{2}+1) \rightarrow (N_{1}+1,N_{2})$, with the total charge  increased only by $1e$. In Fig.~\ref{fig:stabilitydiagrams} we show the stability diagram containing the relevant groundstate sequence. The repulsion $K_{12} > K_{12c}$ needs to have a minimal strength, depending on the precise value of $\phi_{02}$, to stabilize this groundstate sequence. As the coupling $K_{12} $ is further increased for $K_{1} \neq K_{2} $, additional transitions with increasing $ \Delta N_{1} > 2 $ appear in the stability diagram, until the system eventually becomes unstable.
If $K_{12} < K_{12c}$ on the other hand, it is energetically more favourable to place a single additional charge into $N_{1}$ and to leave $N_{2}$ unchanged, instead of adding two charges to $N_{1}$ and redistributing charge afterwards. In the symmetric limit $K_{2} = K_{1}$, $K_{12c} = K_{1}$, and hence the "pairing regime" $K_{12} > K_{1}$ lies in the unstable regime. \\
For the second pairing mechanism in scenario B, the Coulomb repulsion $K_{eb}$ between edges and bulk is the relevant coupling. We assume $K_{1} = K_{2} = K_b$ and $K_{1} > K_{eb} > K_{12} $. As the flux is increased by one flux quantum, the number $N_b$ of bulk particles decreases by $2$, typically in two separate groundstate transitions, while the edge charges increase by $1e$ each. The role of $K_{eb}$ is to `synchronize' these transitions as a function of the flux, such that 
$(N_1, N_2, N_b) \rightarrow (N_1+1, N_2+1, N_b-2)$. The synchronization is eventually destroyed for some finite value of 
$\phi_{02}$, where now the sequence of transitions $(N_1, N_2, N_b) \rightarrow (N_1, N_2+1, N_b-1) \rightarrow (N_1+1, N_2+1, N_b-2)$ takes place. The synchronization turns out to be stable for $K_1 \neq K_2$.
%
\begin{figure}[t!]
	\begin{minipage}{0.45\textwidth}
		\centering
		\flushleft \textbf{Scenario A}
		\\[0.5em]
		\includegraphics[scale=0.35]{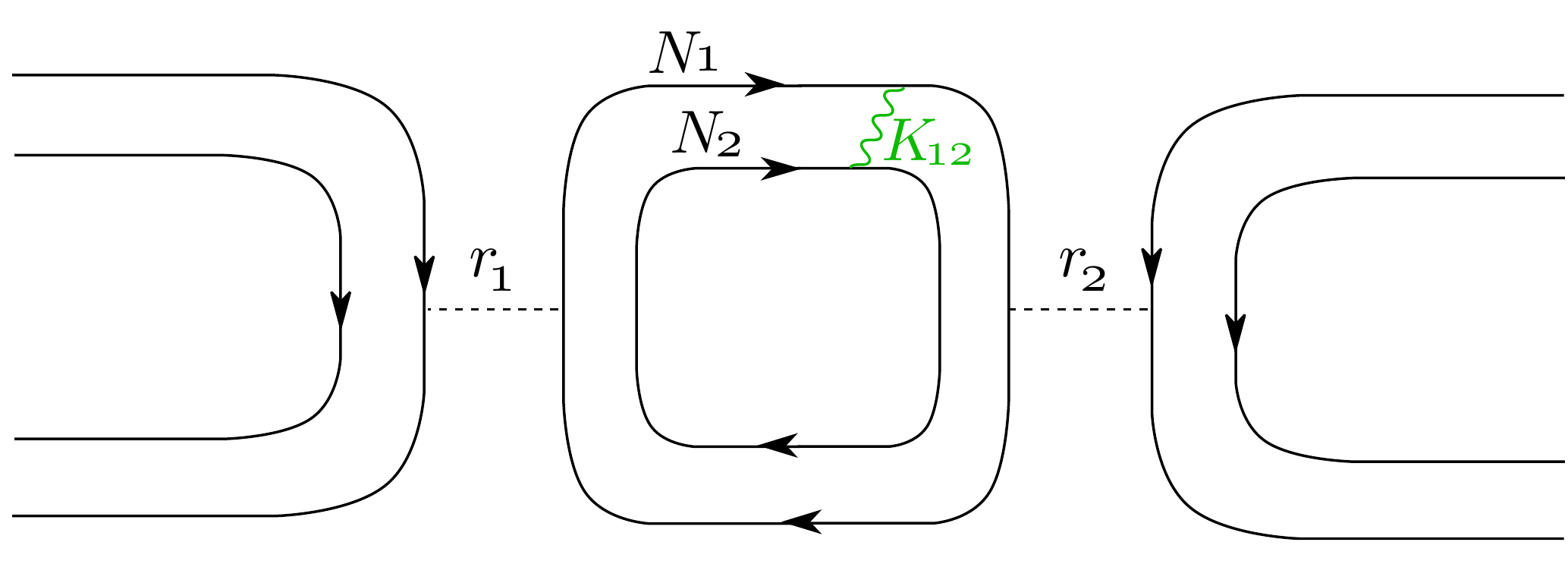}
	\end{minipage}
	\hspace{0.2cm}
	\begin{minipage}{0.45\textwidth}
		\centering
		\vspace{0.2cm}
		\flushleft \textbf{Scenario B}
		\\[0.5em]
		\includegraphics[scale=0.35]{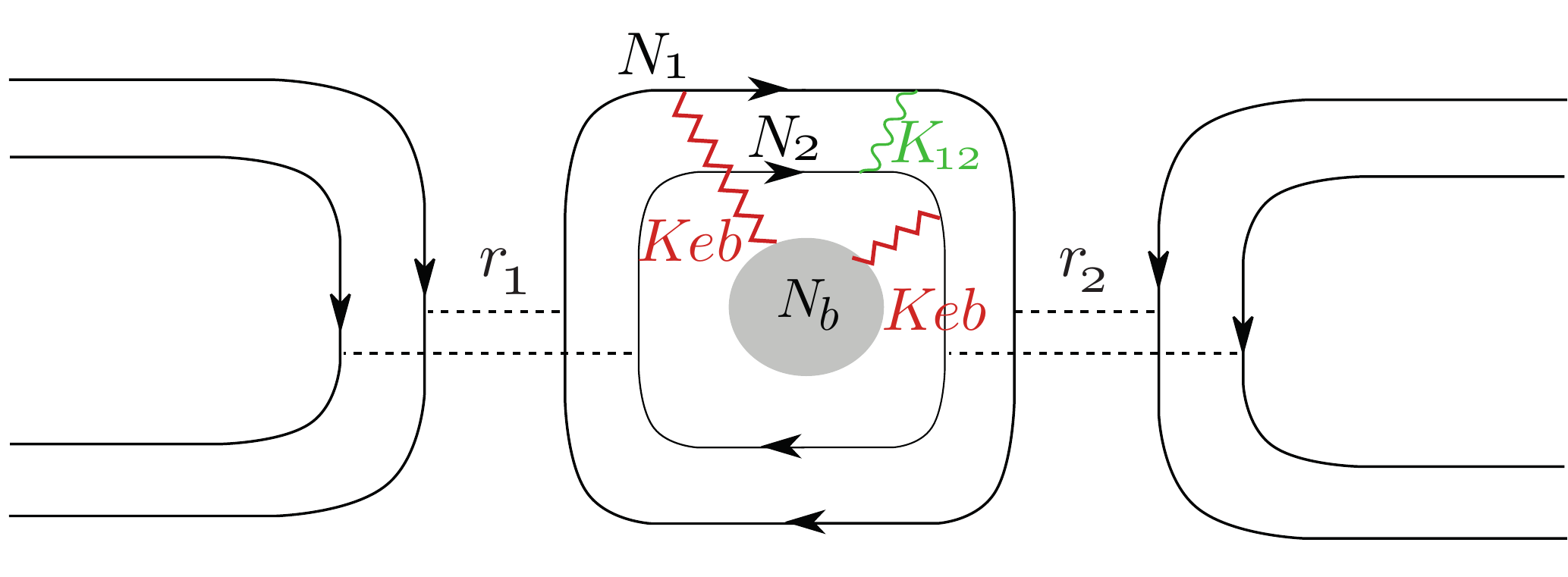}
	\end{minipage}
	\caption{FPI in the closed limit. In scenario A only the outer edge is contacted and the bulk is neglected. On the other side, both outer and inner edges are contacted and the bulk plays an important role in scenario B.}\label{fig:closedFPI}
\end{figure}
%
%
\begin{figure}[t!]
	\begin{minipage}{0.45\textwidth}
		\centering
		\flushleft \textbf{Scenario A}
		\\[0.5em]
		\includegraphics[scale=0.52]{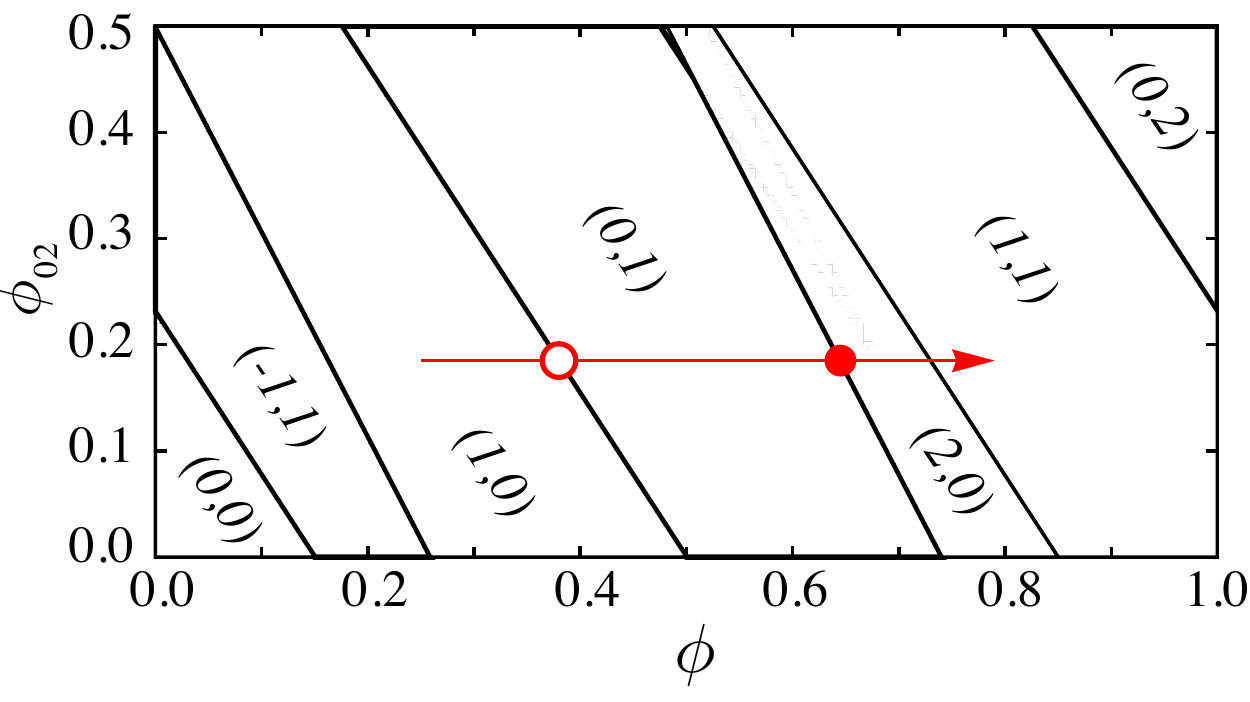}
	\end{minipage}
	\begin{minipage}{0.45\textwidth}
		\centering
		\flushleft \textbf{Scenario B}
		\\[0.5em]
		\includegraphics[scale=0.52]{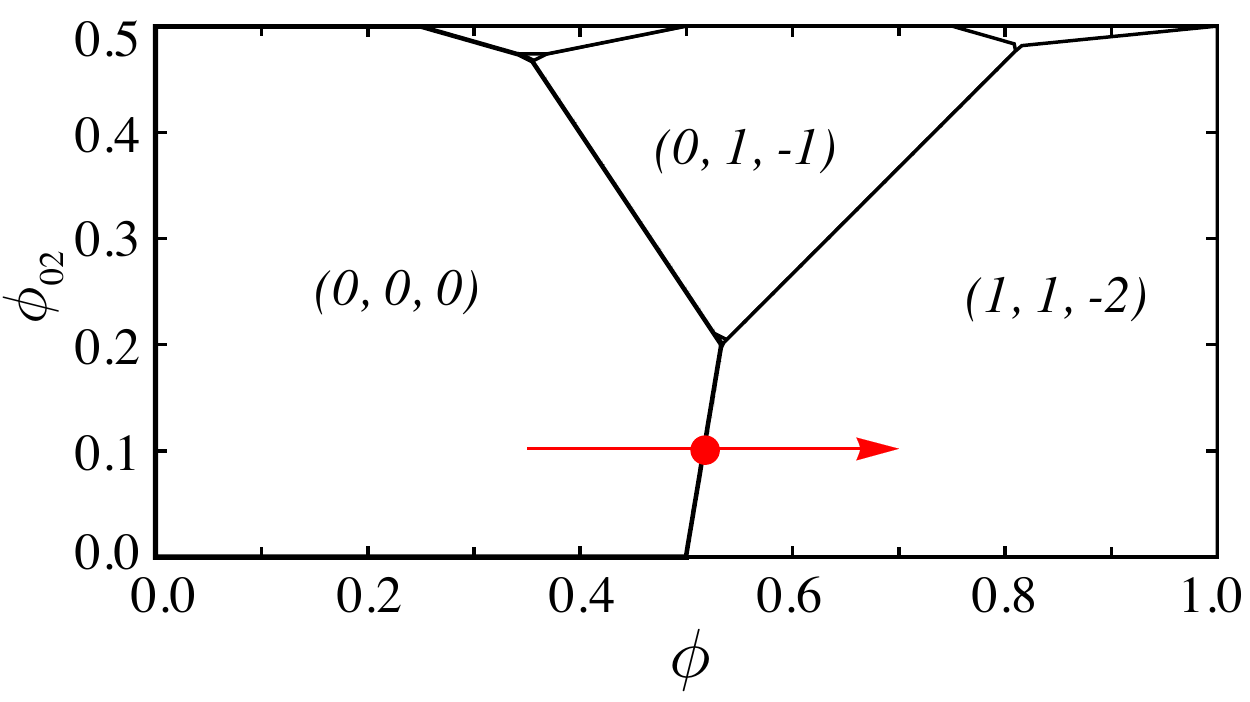}
	\end{minipage}
	\caption{Groundstate stability diagrams as a function of flux $\phi$ and the offset parameter  $\phi_{02}$ for the closed interferometer. The groundstates are differentiated by their charge configuration $(N_{1},N_{2})$ in the inner and outer edge, respectively. (a) Corresponding to the pairing scenario A, we take $ K_{2} = 2 K_{1} $ and $ \Lambda = 1.35 $. The red arrow marks a flux-driven sequence of groundstate transitions that contains a charge redistribution between the edges (open red circle) and a subsequent transition accompanied by $\Delta N_{1} = 2$ (red disk). (b) The stability diagram for the pairing scenario B with the charge configurations $(N_{1},N_{2},N_{b})$, where we set $K_{1} = K_{2} = K_{b}$, $\Delta = 0.8$ and $\Lambda = 0.5$. The red disk denots the flux driven charge-redistribution transition involving a simultaneous change of $N_{1}$ and $N_{2}$ by 1 each, while the bulk charge changes by -2.}
	\label{fig:stabilitydiagrams}
\end{figure}
%
\section{\label{sec:Conductance in the closed limit} S5. Conductance in the closed limit}
\begin{figure}
	\centering
	\begin{minipage}{0.24\textwidth}
		\begin{flushleft}\hspace{0.cm}
			\includegraphics[scale=0.35]{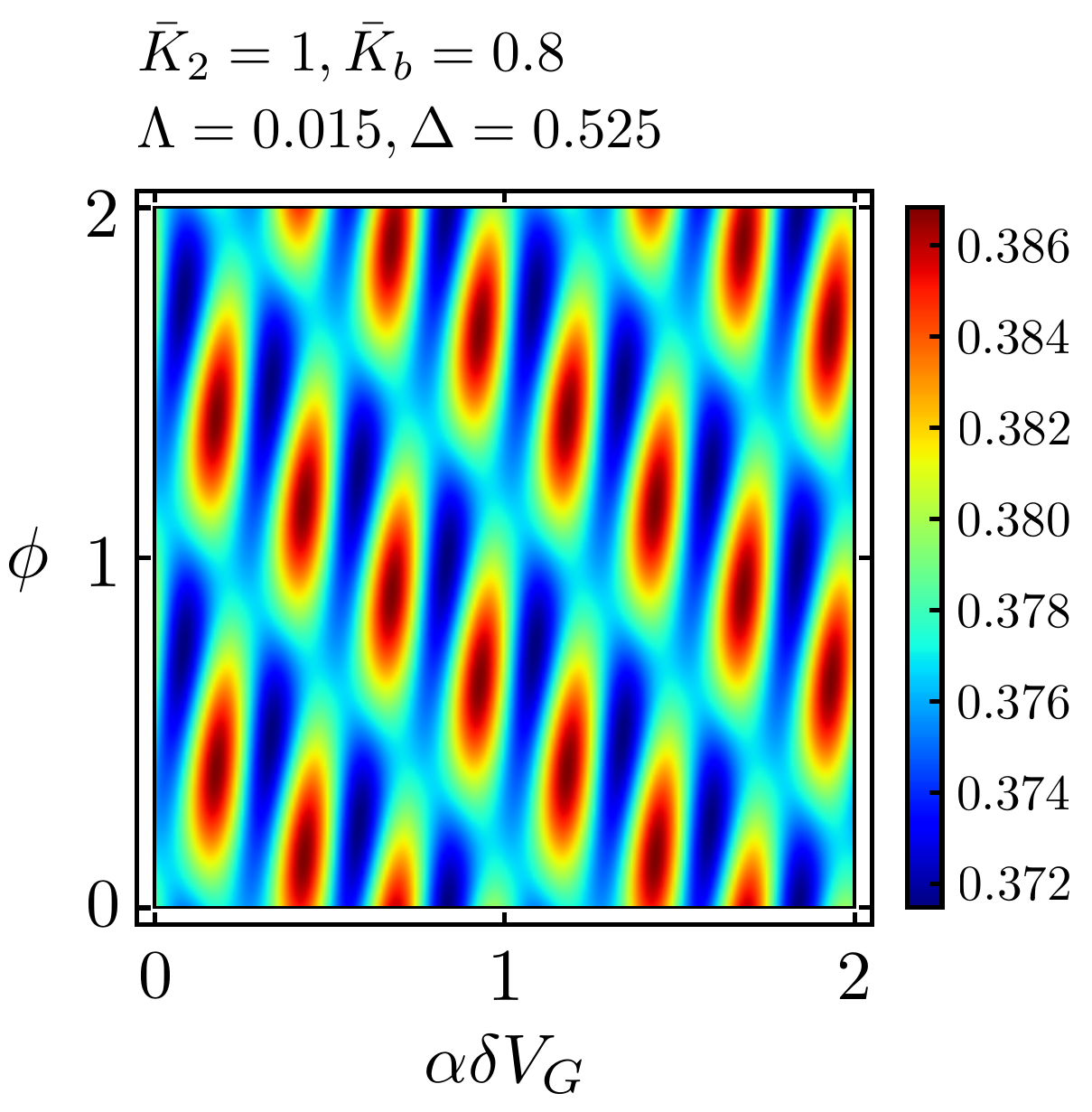}
		\end{flushleft}
	\end{minipage}
	\begin{minipage}{0.24\textwidth}
		\begin{flushleft}\hspace{0.cm}
			\includegraphics[scale=0.35]{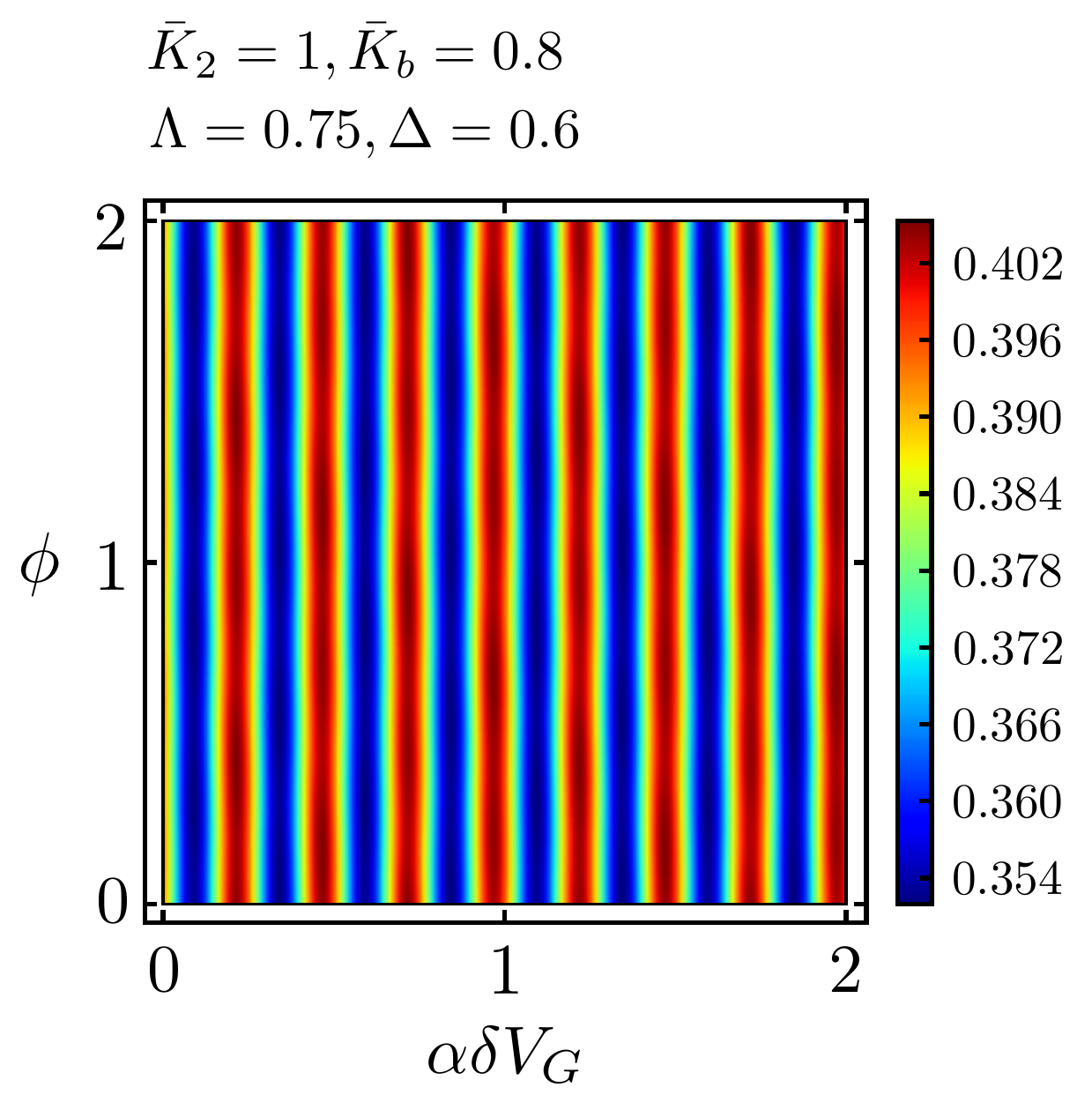}
		\end{flushleft}
	\end{minipage}
	\begin{minipage}{0.24\textwidth}
		\begin{flushleft}\hspace{0.cm}
			\includegraphics[scale=0.35]{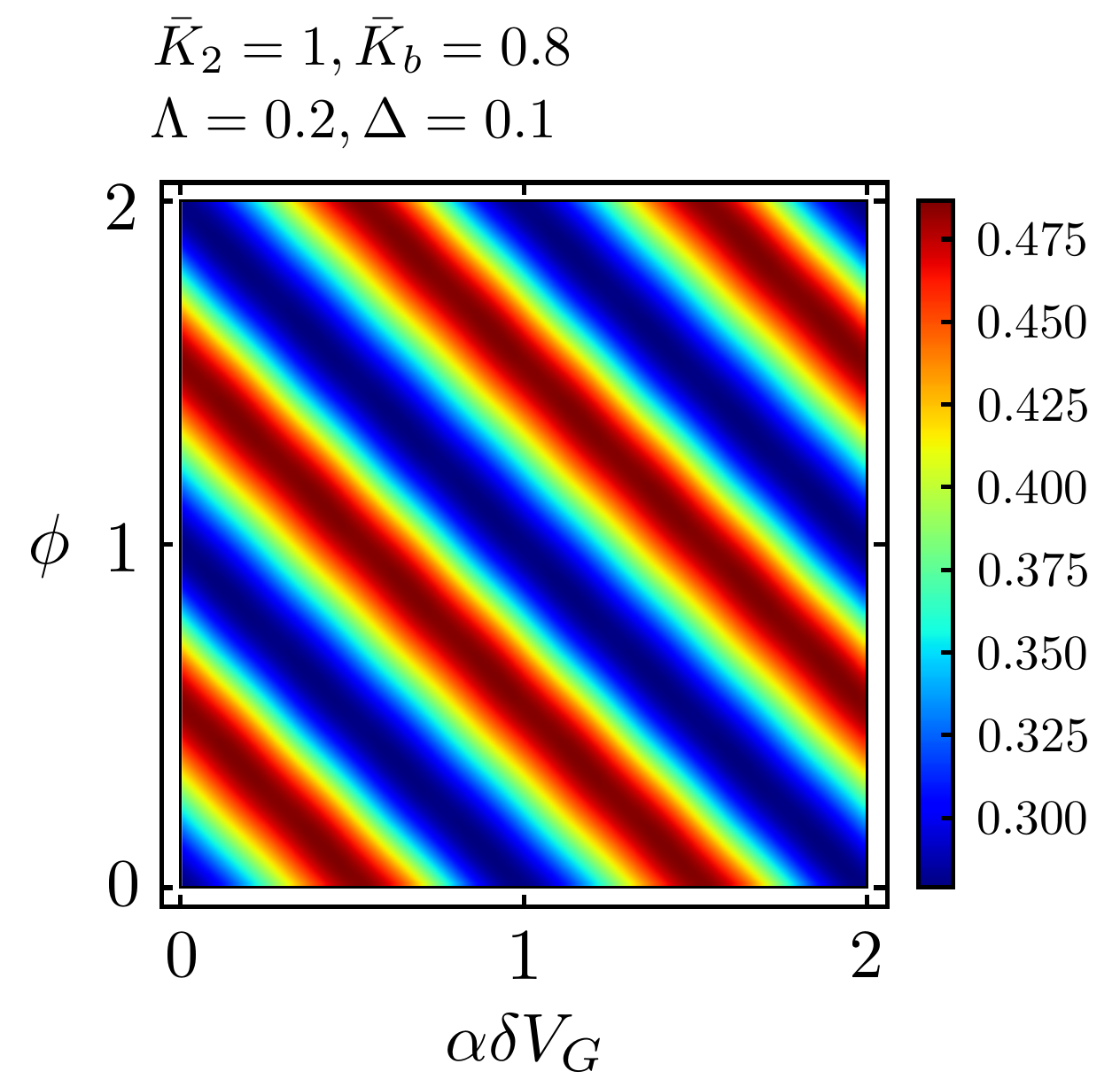}
		\end{flushleft}
	\end{minipage}
	\begin{minipage}{0.24\textwidth}
		\begin{flushleft}\hspace{0.cm}
			\includegraphics[scale=0.35]{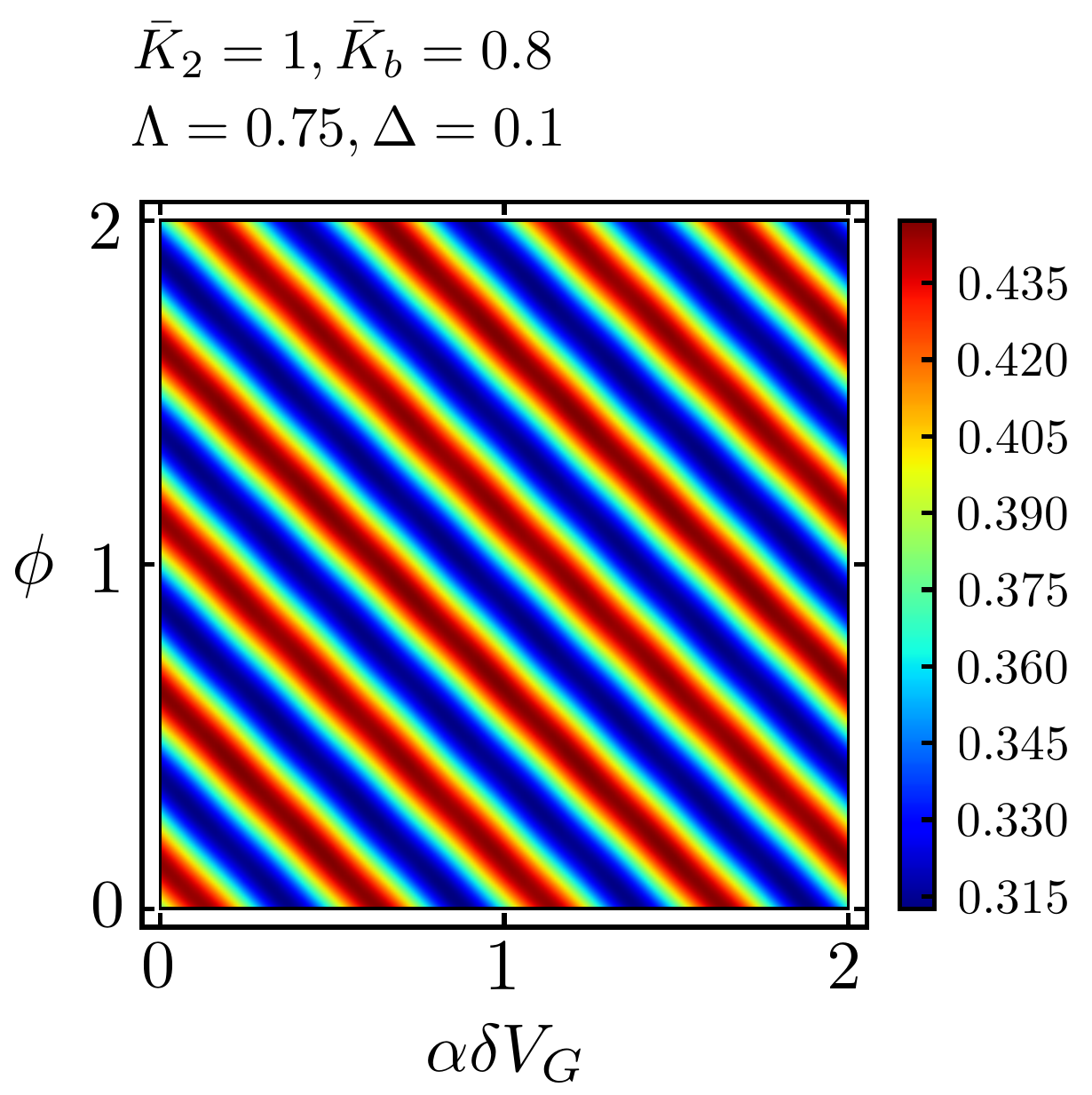}
		\end{flushleft}
	\end{minipage}
	\caption{Conductance in the closed limit as a function of $\alpha\delta V_g$ and $\phi$ according to \Eqref{eq:diffusion rate} for the four different regions of the phase diagram. We set $\gamma=4\alpha, \phi_{02}=0.25, \phi_{0b}=0.3$ and $\beta=5$.}
	\label{fig:conductance closed limit}
\end{figure}

The conductance in the closed limit can be calculated via the fluctuation-dissipation theorem, $\delta G=M \beta \tilde{D}$, where $\tilde{D}$ is the diffusion rate for motion of electrons inside or outside the interfering edge \cite{s-PhysRevLett.98.106801}
%
\begin{equation}\label{eq:diffusion rate}
\tilde{D}=\frac{\sum_{N_1,N_2,N_b} e^{-\beta E(N_1,N_2,N_b)}\left[f(\Delta_+^1)+1-f(-\Delta_-^1)\right]}{\sum_{N_1,N_2,N_b} e^{-\beta E(N_1,N_2,N_b)}},   \qquad \Delta_\pm^1=E(N_1\pm 1,N_2,N_b)-E(N_1,N_2,N_b)
\end{equation}
%
with $f(x)=(1+e^{\beta x})^{-1}$ denoting the Fermi distribution, $M$ contains the tunneling matrix elements which we assume to be constant for simplicity and $\Delta_\pm^1$ are the addiction/subtraction energies of one electron from the interfering edge mode. 
%
%
We present in Fig.~\ref{fig:conductance closed limit} the density plot of the conductance as a function of $\alpha \delta V_G$ and $\phi$ obtained from \Eqref{eq:diffusion rate} for different values of the couplings. We note that the slope of the lines of constant phase and the periodicities of the conductance coincide with the one calculated in the open limit. Therefore, the results achieved in the open limit are valid also for the closed limit. Moreover, the conductance is maxim/minimum when $\text{min}(\Delta_+^1,\Delta_-^1)$, evaluated in the ground state configuration, is minimum/maxim, as it can be seen from comparison of Fig.~\ref{fig:conductance closed limit} with Fig.~(4) in the main text. This is something expected, indeed the conductance is big/small when we need a small/big amount of energy for adding, or subtracting, an electron.
%
%

%

\end{document}